\documentclass[aps,letterpaper,twocolumn,prl,showpacs,nofootinbib]{revtex4-1}
\usepackage{amssymb}
\usepackage{amsmath,bm}

\usepackage[pdftex]{graphicx,color}
\usepackage[english]{babel}
\usepackage{hyperref}
\usepackage{enumitem}
\usepackage{epstopdf}
\usepackage{tikz}

\begin{document}

\title{Experimental observations of bright dissipative Kerr cavity solitons and their collapsed snaking in a driven resonator with normal dispersion}

\author{Zongda Li$^{1,2}$}
\author{St\'ephane Coen$^{1,2}$}
\author{Stuart G. Murdoch$^{1,2}$}
\author{Miro Erkintalo$^{1,2,}$}
\email{m.erkintalo@auckland.ac.nz}
\affiliation{$^1$Department of Physics, University of Auckland, Auckland 1010, New Zealand}
\affiliation{$^2$The Dodd-Walls Centre for Photonic and Quantum Technologies, New Zealand}

\begin{abstract}
Driven Kerr nonlinear optical resonators can sustain localized structures known as dissipative Kerr cavity solitons, which have recently attracted significant attention as the temporal counterparts of microresonator optical frequency combs. Whilst conventional wisdom asserts that bright cavity solitons can only exist in the region of anomalous dispersion, recent theoretical studies have predicted that higher-order dispersion can fundamentally alter the situation, enabling bright localized structures even under conditions of normal dispersion driving. Here we demonstrate a flexible optical fibre ring resonator platform that offers unprecedented control over dispersion conditions, and we report on the first experimental observations of bright localized structures that are fundamentally enabled by higher-order dispersion. In broad agreement with past theoretical predictions, we find that several distinct bright structures can co-exist for the same parameters, and we observe experimental evidence of their collapsed snaking bifurcation structure. In addition to enabling direct experimental verifications of a number of theoretical predictions, we show that the ability to judiciously control the dispersion conditions offers a novel route for ultrashort pulse generation: the bright structures observed in our work correspond to pulses of light as short as 220~fs -- the record for a passive all-fibre ring resonator. We envisage that our work will stimulate further fundamental studies on the impact of higher-order dispersion on Kerr cavity dynamics, as well as guide the development of novel ultrashort pulse sources and dispersion-engineered microresonator frequency combs.

\end{abstract}

\maketitle

\section{Introduction}
\noindent In 1983, McLaughlin, Moloney, and Newell predicted the existence of transverse localized structures in a passive, coherently-driven nonlinear optical cavity~\cite{laughlin_solitary_1983}. The subsequent two decades saw the dynamics and characteristics of such structures -- today known as spatial cavity solitons (CSs) -- extensively investigated, motivated by fundamental interest and application prospects alike~\cite{firth_optical_1996, spinelli_spatial_1998, firth_cavity_2002, ackemann_chapter_2009}. These studies arguably culminated in the pioneering experiments of Barland et al., who in 2002 showed that a semiconductor microcavity could sustain CSs in the form of transversely self-localized (non-diffracting) beams that could be turned on and off at will~\cite{barland_cavity_2002}.

In 1993, Wabnitz used the analogy between spatial diffraction and temporal dispersion to predict the existence of temporal analogs of spatial CSs in passive, coherently-driven Kerr nonlinear ring resonators~\cite{wabnitz_suppression_1993}. Such temporal structures correspond to ``bright'' pulses of light that sit atop a low-intensity continuous wave background, and that can circulate around the resonator indefinitely, without changing their shape or energy. They attracted scant attention until 2010, when they were first experimentally observed by Leo et al. in a macroscopic fibre ring resonator~\cite{leo_temporal_2010}. In homage to the extensive and pioneering studies performed in the context of spatially diffractive resonators, the authors coined the term temporal CSs to describe the structures. Following the first experimental observation of temporal CSs, numerous studies embraced the unprecedented degree of control afforded by fibre-based systems to explore fundamental CS physics: experimental evidence of oscillatory (Hopf) instabilities~\cite{leo_dynamics_2013}, emission of dispersive waves~\cite{jang_observation_2014}, control with phase gradients~\cite{jang_temporal_2015,jang_all-optical_2016}, coexistence of distinct CS states~\cite{anderson_coexistence_2017, nielsen_coexistence_2019}, and a number of other phenomena~\cite{jang_ultraweak_2013, anderson_observations_2016, xue_super-efficient_2019} have been observed over the past decade.

In 2013, Herr et al. reported on the experimental observation of temporal CSs in a monolithic Kerr microresonator -- a \emph{microscopic} analogue of a \emph{macroscopic} fibre ring resonator~\cite{herr_temporal_2014}. It is in such microresonators that the full application potential of (temporal) CSs has finally been unleashed -- more than three decades after studies on localized structures in passive cavities first emerged~\cite{laughlin_solitary_1983}. Specifically, microresonator CSs (also known as dissipative Kerr solitons) correspond to coherent and (potentially) broadband optical frequency combs~\cite{delhaye_optical_2007, brasch_photonic_2016, obrzud_temporal_2017, cole_kerr-microresonator_2018, lucas_spatial_2018, jang_synchronization_2018, kippenberg_dissipative_2018}, whose utility has now been demonstrated in a host of applications, including telecommunications~\cite{marin-palomo_microresonator-based_2017}, optical distance measurements~\cite{trocha_ultrafast_2018,suh_soliton_2018}, spectroscopy~\cite{suh_microresonator_2016, dutt_-chip_2018}, and optical frequency synthesis~\cite{spencer_optical-frequency_2018}. Moreover, the unique characteristics of microresonators (e.g. strong thermal nonlinearity~\cite{carmon_dynamical_2004} and complex mode interactions~\cite{herr_mode_2014}) has been shown to engender rich new physics, such as the emergence of periodic soliton crystals~\cite{cole_soliton_2017, karpov_dynamics_2019}, soliton switching~\cite{guo_universal_2016}, and single-mode dispersive waves~\cite{yi_single-mode_2017}.

It is commonly held that ``bright'' CSs -- intensity peaks atop a low-level background -- only manifest themselves under conditions of a self-focusing nonlinearity, which for common resonator configurations (with a positive Kerr nonlinearity coefficient) implies anomalous group velocity dispersion. In stark contrast, the normal dispersion regime can permit ``dark'' solitons -- intensity dips on a high-level background -- which were first observed in a Kerr resonator by Xue et al. in 2015~\cite{xue_mode-locked_2015}. Such dark solitons have been explained to arise through a fundamentally different mechanism than their counterparts in the anomalous dispersion regime~\cite{parra-rivas_origin_2016}. Specifically, whereas the latter can be understood as a single cycle of an underlying Turing (or modulation instability, MI) pattern, the former arise via the interlocking of switching waves (or fronts): nonlinear structures that connect the two homogeneous states of the bistable cavity system. This fundamental difference gives dark solitons particular advantages, such as enhanced spectral conversion efficiency~\cite{xue_microresonator_2017}, which has recently been leveraged for high-order coherent communications~\cite{fulop_high-order_2018, helgason_superchannel_2019}.

Whilst bright CSs are not, under typical conditions, to be expected in the normal dispersion regime, studies have shown that higher-order perturbations can fundamentally alter the situation. For example, interactions between different transverse mode families can  permit bright localized structures with flat-top profiles (``platicons'')~\cite{lobanov_frequency_2015, lobanov_generation_2015, jang_dynamics_2016, lobanov_dynamics_2017}, while frequency-dependent cavity losses can similarly give rise to a variety of bright soliton structures in the normal dispersion regime~\cite{huang_mode-locked_2015, spiess_chirped_2019}. Of particular significance is the recent theoretical prediction made by Parra-Rivas et. al. that even the simplest of perturbations -- third-order dispersion (TOD) -- is sufficient to allow bright structures in Kerr cavities with normal dispersion~\cite{parra-rivas_coexistence_2017}. Like dark solitons, such TOD-enabled bright CSs are predicted to arise via interlocking of switching waves, and for a given system parameters, several of them can co-exist, arranged in a bifurcation structure referred to as collapsed snaking.

To the best of our knowledge, no experimental observations of TOD-enabled CSs have yet been reported. This is arguably because existing experimental configurations are not suitable for systematic studies of TOD effects in isolation. Microresonator devices are hindered by thermal nonlinearities~\cite{carmon_dynamical_2004} and mode interactions~\cite{herr_mode_2014}, whilst fiber ring resonator systems typically operate under conditions where TOD is negligible~\cite{leo_temporal_2010,leo_dynamics_2013, anderson_coexistence_2017, nielsen_coexistence_2019, jang_ultraweak_2013, anderson_observations_2016, xue_super-efficient_2019}. A handful of studies have used dispersion-managed fibre resonators to enhance the impact of TOD~\cite{jang_observation_2014,copie_competing_2016,copie_dynamics_2017, wang_universal_2017}, but in this case the dispersion-management itself constitutes a significant perturbation that can obfuscate the cavity dynamics~\cite{nielsen_invited_2018}.

In this Article, we report on a flexible experimental platform that permits systematic studies of Kerr cavity dynamics in the presence of higher-order dispersion. Our system comprises of a homogeneous (i.e., non-dispersion-managed) ring resonator made entirely out of dispersion shifted fibre with a zero-dispersion wavelength (ZDW) at 1565~nm; by driving the resonator with a widely-tunable external cavity diode laser, we are able to controllably explore dynamics around the ZDW, and hence control the impact of higher-order dispersion. We observe clear evidence of bright CSs that are fundamentally enabled by TOD, and we identify experimental signatures of their collapsed snaking bifurcation structure. Our experiments also reveal that the range of existence of the conventional CSs that manifest themselves in the anomalous dispersion regime can extend into the normal dispersion regime~\cite{milian_soliton_2014}, and we discuss the relationship between the localized structures appearing in the different dispersion regimes. The ability to examine CS characteristics across the ZDW has also allowed us to observe -- for the first time to the best of our knowledge -- the emission of spectrally symmetric dispersive waves as predicted in earlier theoretical works~\cite{milian_soliton_2014}. Taken together, our experiments confirm a string of past theoretical predictions, and they unveil a rich range of novel dynamics in Kerr resonators operating close to the ZDW. From an applied perspective, the prospect of systematically generating localized structures in the normal dispersion regime (and arbitrarily close to the ZDW) could pave the way for novel sources of ultrashort pulses and broadband optical frequency combs. In this context, with durations as short as 220~fs, the bright structures observed in our experiments correspond -- to the best of our knowledge -- to the shortest CSs generated in a passive fibre ring resonator.

\section{Results}

\subsection{Experimental setup}

\begin{figure*}[!t]
 \centering
  \includegraphics[width = \textwidth, clip=true]{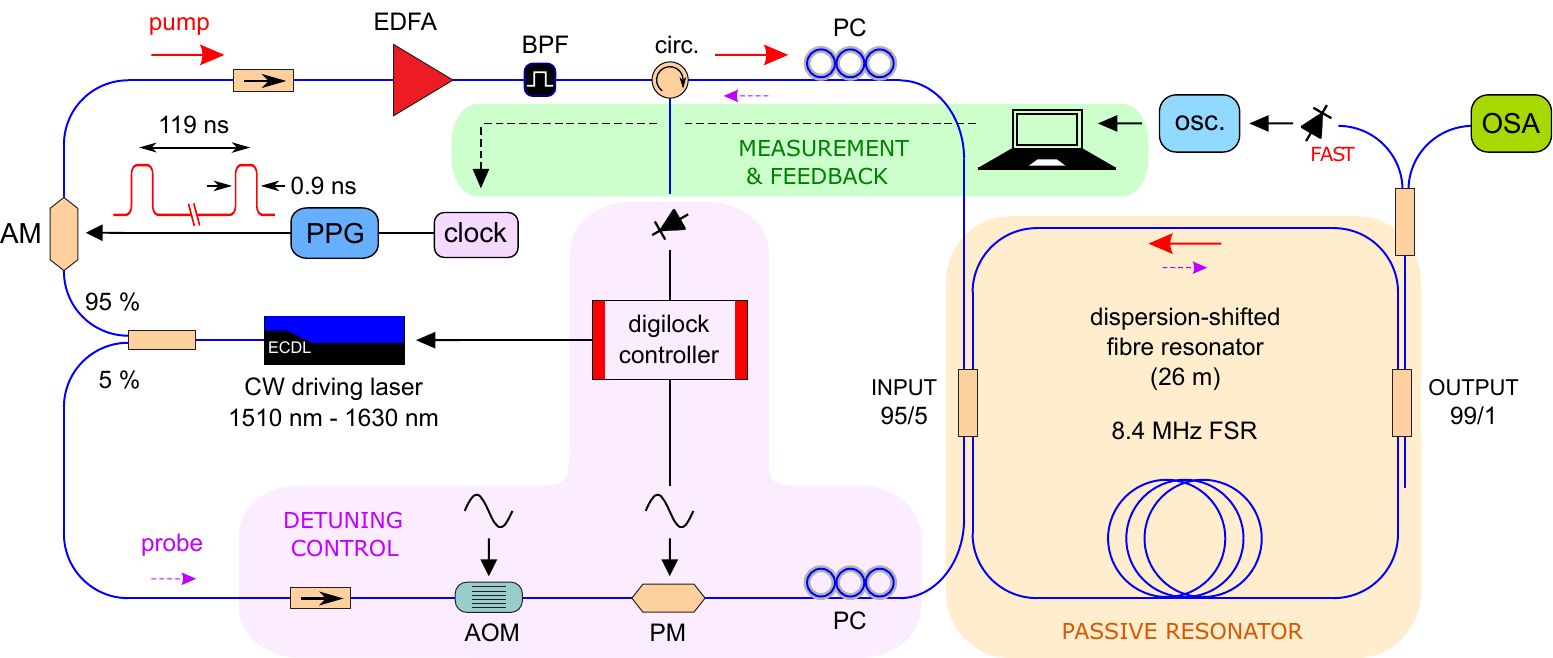}
 \caption{Schematic illustration of the experimental setup. A passive fibre ring resonator made out of a 26-m-long segment of dispersion-shifted fibre [orange shaded region] is driven with nanosecond pulses carved from an external-cavity diode laser (ECDL). The pump repetition rate is actively locked to the intracavity soliton repetition rate by a computer-based measurement \& feedback system [green shaded region]. The detuning between the ECDL frequency and a linear cavity resonance is actively stabilized using the Pound-Drever-Hall (PDH) technique [magenta shaded region]; a digital laser locking module (Toptica Digilock) simultaneously provides the PDH modulation signal, demodulates the photodetector signal measured at the cavity output, and acts as the proportional-integral-derivative controller for the driving laser. AM, amplitude modulator; PPG, pulse-pattern generator; EDFA, Erbium-doped fibre amplifier; BPF, band-pass filter; circ., circulator; PC, polarization controller; AOM, acousto-optic modulator; PM, phase modulator; OSA, optical spectrum analyzer; osc., oscilloscope.}
 \label{Fig1}
 \vskip-10pt
\end{figure*}

\noindent Figure~\ref{Fig1} shows a schematic illustration of our experimental platform [see also Methods]. At the heart of it lies a passive fibre ring resonator made of a 26~m long segment of dispersion shifted fibre (DSF). The resonator incorporates two couplers made out of the same DSF as the main cavity: a first coupler with 95/5 coupling ratio is used to inject the driving field into the cavity, whilst a second coupler with 99/1 coupling ratio is used to extract a small portion (1\%) of the intracavity field for analysis. The cavity has a free-spectral range (FSR) of 8.1~MHz and a measured finesse of $\mathcal{F} = 45$, corresponding to 180~kHz resonance linewidth.

We drive the resonator with flat-top nanosecond pulses carved from an external cavity diode laser (ECDL) whose wavelength can be continuously tuned from 1510~nm to 1630~nm (Toptica CTL). An Erbium-doped fibre amplifier is used to amplify the driving field, thus allowing for the peak power of the nanosecond pulses to be adjusted over a wide range. An external clock signal generated by an RF signal generator is used to carefully adjust the repetition rate of the nanosecond pulse train to match the round trip time of the solitons in the resonator; a computer-automated measurement \& feedback scheme [see green shaded area in Fig.~\ref{Fig1}] continuously monitors and corrects the clock signal so as to maintain the solitons close to the centre of the driving pulse. In this context, we must emphasize that the nanosecond duration of our driving pulses is considerably longer than the sub-picosecond durations of the solitons under study. As a consequence, the solitons experience the driving field effectively as continuous wave.

\begin{figure*}[!t]
 \centering
  \includegraphics[width = \textwidth, clip=true]{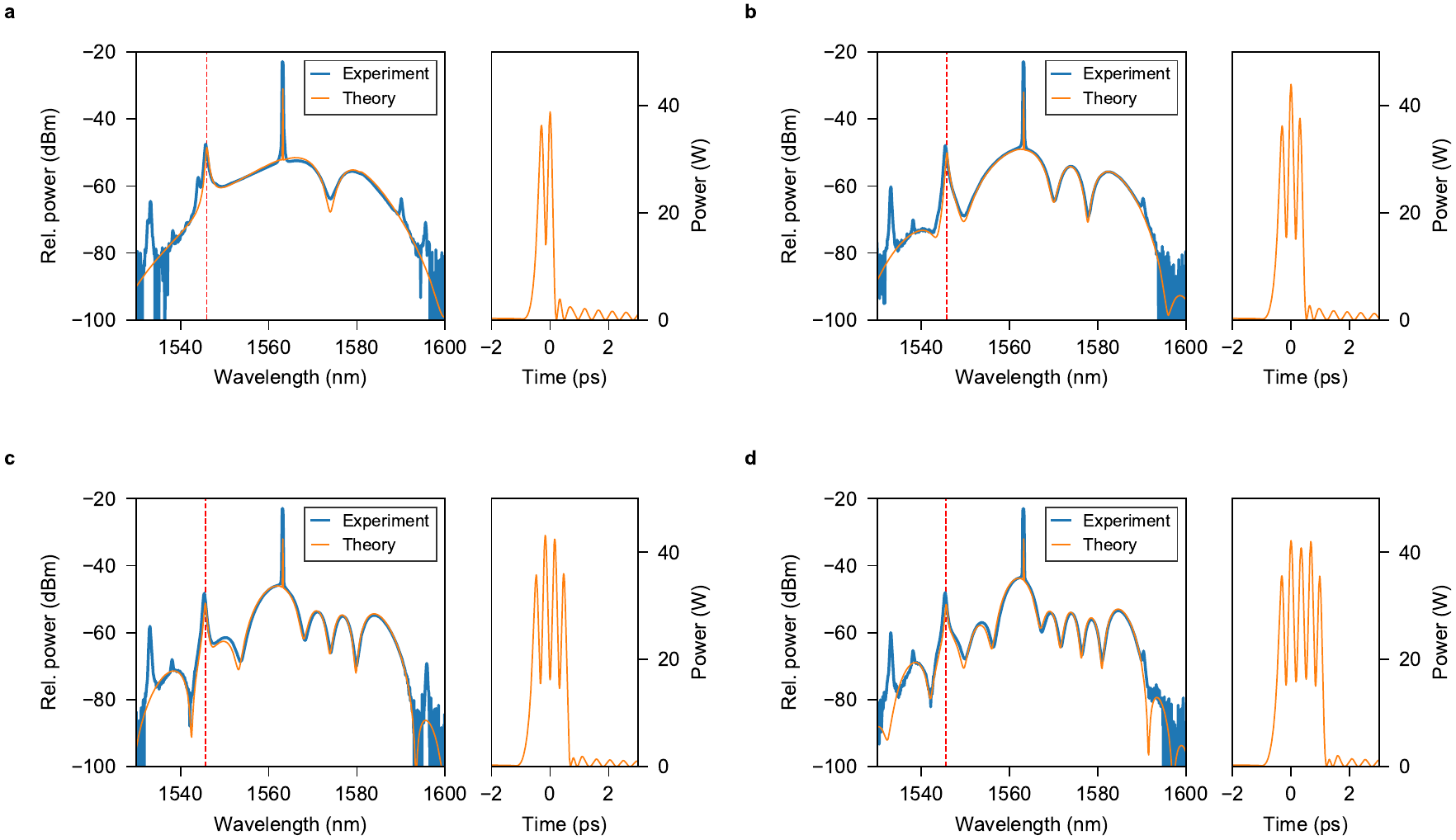}
 \caption{Observations of different bright structures with normal dispersion driving at 1563~nm. \textbf{a}--\textbf{d} The left panels show experimentally measured (blue solid curves) and numerically simulated (orange solid curves) optical spectra, whilst the right panels show corresponding temporal profiles extracted from numerical simulations. Dashed vertical lines highlight the theoretically predicted dispersive wave wavelengths (see Methods). All results obtained for the same system parameters, including pump peak power (6.7~W) and linear cavity detuning ($\delta_0 = 1.05~\mathrm{rad}$).}
 \label{fig2}
 \vskip-10pt
\end{figure*}

To stabilize the linear phase detuning \mbox{$\delta_0 = 2\pi(f_0-f_\mathrm{P})/\text{FSR}$} between the driving laser (with carrier frequency $f_\mathrm{P}$) and a cavity resonance (with frequency $f_0$), we launch a low-power cw beam derived from the ECDL into the cavity such that it counter-propagates with respect to the main pump, and use the Pound-Drever-Hall (PDH) technique to lock the laser frequency to the peak of the probe's Lorentzian resonance [magenta shaded region in Fig.~\ref{Fig1}]. By using an acousto-optic modulator (AOM) to frequency shift the probe beam before its injection into the cavity, we are able to continuously tune the linear detuning experienced by the main pump~\cite{nielsen_invited_2018}.

Because dispersion plays a central role in our study, we have carefully characterised the dispersive properties of our cavity. This was achieved by measuring, for a wide range of pump wavelengths, the frequency shifts of spectral sidebands generated via phase-matched four-wave mixing with the detuning stabilised at zero (see Supplementary Information). By fitting a theoretically predicted phase-matching curve on the experimentally measured tuning curve, we find that the cavity has a ZDW of 1565.4~nm and third- and fourth-order dispersion coefficients of $\beta_{3,\text{ZDW}} = 0.135~\mathrm{ps^3/km}$ and \mbox{$\beta_{4,\text{ZDW}} = -9\times10^{-4}~\mathrm{ps^4/km}$} at the ZDW, respectively. In our experiments, we set the pump wavelength close to the ZDW, thus enhancing the impact of TOD [see Methods]. This should be contrasted with an earlier study by Bessin et al., who used a uniform resonator made out of DSF to study the dynamics of modulation instability in the normal dispersion regime~\cite{bessin_modulation_2017}; no signatures of bright CSs were reported since the resonator was driven far from the ZDW where third-order dispersion is negligible.

Our experimental platform incorporates a number of advances compared to earlier studies on CSs, each crucial to obtaining the results presented in the following sections. First, the use of a homogeneous cavity with no dispersion management ensures that effects observed are representative of pure Kerr cavity physics. Second, the use of a widely-tunable ECDL allows us to systematically explore the cavity dynamics in the vicinity of the ZDW, providing control over the sign of the group-velocity dispersion as well as the relative magnitude of higher-order dispersion. Third, in contrast to active stabilization schemes used in prior studies, our PDH-based technique permits the cavity detuning to be linearly and continuously tuned over a wide range, and together with our computer-automated measurement \& feedback scheme, allows for the robust study of detuning-dependent soliton dynamics. Our experiments also suggest that the PDH-based scheme is more efficient than the side-of-fringe locking applied in previous studies. In fact, we find that a simple side-of-fringe lock used e.g. in~\cite{nielsen_invited_2018} is not sufficient to allow the pump ECDL, with a measured linewidth of 80~kHz [see Methods], to coherently drive our cavity; in contrast, the PDH scheme allows the detuning to be more robustly stabilized, enabling us to observe and study coherent cavity dynamics over a wide range of parameters.

\subsection{First observations of TOD-enabled bright CSs}

\begin{figure*}[!t]
 \centering
  \includegraphics[width = \textwidth, clip=true]{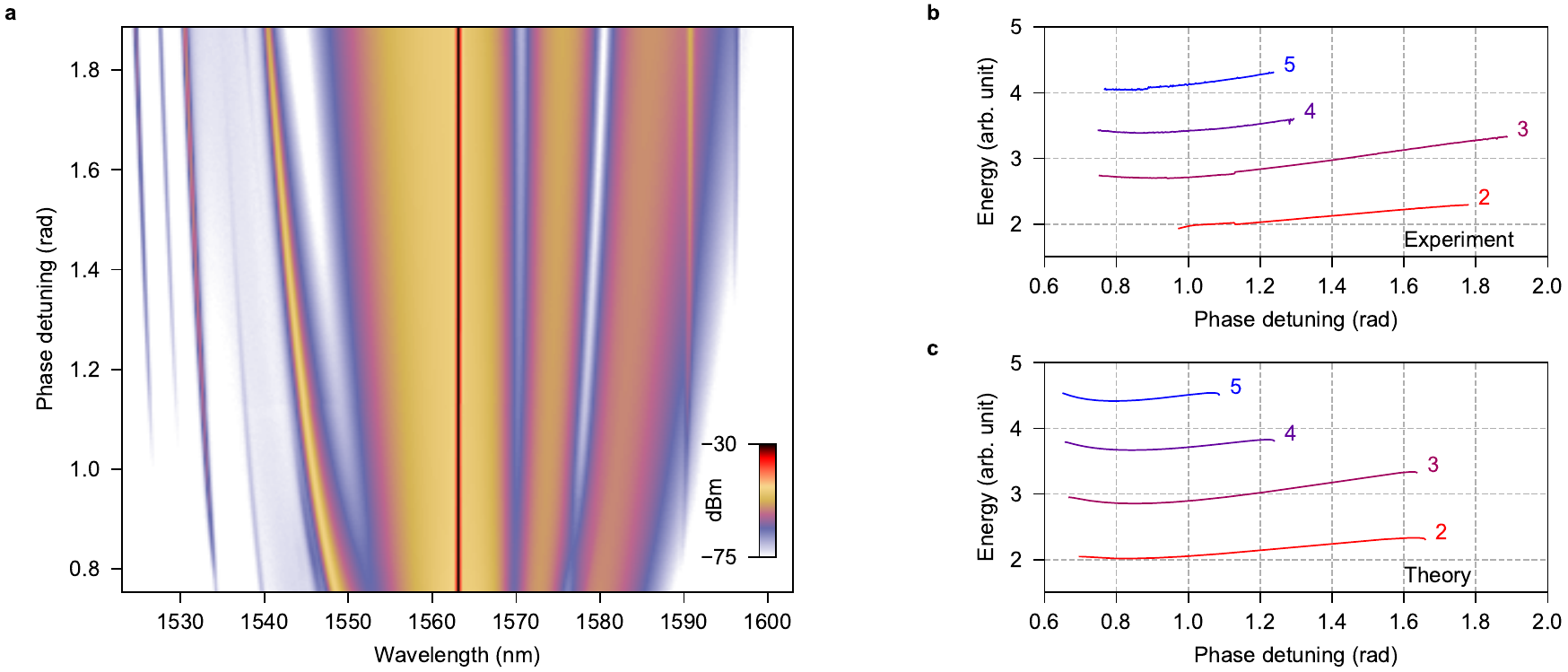}
 \caption{Signatures of collapsed snaking. \textbf{a,} 704 optical spectra measured at different detunings for a three-peak soliton, concatenated vertically to form a single pseudo-color plot. \textbf{b,} Experimentally measured bifurcation curve for the different bright structures considered in Fig.~\ref{fig2}. \textbf{c,} Corresponding theoretical bifurcation curve. Numbers next to the curves correspond to the number of peaks the different structures possess in their temporal profile. }
 \label{fig3}
 \vskip-10pt
\end{figure*}

\noindent We first present illustrative experimental observations of bright CSs in the normal dispersion regime. The results described below were obtained with the pump wavelength and power set to 1563~nm (2~nm below the ZDW) and 6.7~W (peak power of the flat-top nanosecond pulses), respectively, yet we note that similar observations have been obtained for a range of parameter values.

Because our fibre cavity exhibits a negative fourth-order dispersion coefficient $\beta_4$, the upper state of the bistable cavity response is modulationally unstable even in the normal dispersion regime~\cite{bessin_modulation_2017, sayson_octave-spanning_2019}. As a consequence, it is not possible to sustain dark solitons in our system. In stark contrast, we find that a negative $\beta_4$ does not prevent or even noticeably perturb the bright CSs that are fundamentally enabled by TOD; rather, the MI permitted by fourth-order dispersion in fact offers a convenient route for the solitons to be spontaneously excited simply by scanning the detuning over a cavity resonance. We must emphasize, however, that the soliton existence does not require a negative $\beta_4$ (or MI), nor do the solitons correspond to localized elements of any $\beta_4$-induced MI pattern; the MI only facilitates the solitons' spontaneous formation [see Supplementary Information].

In our experiments, we lock the detuning close to the point where the system becomes bistable (as identified by the collapse of the MI state to the lower homogeneous state). We then mechanically perturb the cavity, thus causing the detuning stabilization system to scan across a resonance; solitons emerge spontaneously as a result. In agreement with theoretical predictions~\cite{parra-rivas_coexistence_2017}, we find that several soliton states with distinct spectral and temporal profiles can coexist for the same cavity detuning. The solid blue curves in Figs.~\ref{fig2}(a)--(d) show the optical spectra measured for four such states, obtained by repeatedly perturbing the cavity with the detuning locked at $1.05~\mathrm{rad}$. The orange curves show corresponding results from numerical simulations of a generalized mean-field model (see Supplementary Information for details regarding the model) with no free-running parameters.

The experimentally measured and numerically simulated spectra in Fig.~\ref{fig2} show very good overall agreement. The only discernible discrepancy is the presence of Kelly-like sidebands in the experimental measurements~\cite{nielsen_invited_2018}; these are not reproduced by a mean field model of the cavity but arise when considering a more general lumped cavity model (not shown). Our simulations further show that the spectra shown in Fig.~\ref{fig2} correspond to bright structures with complex multi-peak profiles in the time domain. Whilst limited detection bandwidth prohibits us from experimentally resolving the structures' sub-picosecond temporal features, photodetector measurements at the 1\% cavity output port provide unequivocal evidence that our observations correspond to bright solitons [see Supplementary Information]. As such, these results constitute unequivocal evidence of TOD-enabled bright CSs under conditions of normal dispersion driving.

It is interesting to note that, whilst each of the four spectra shown in Fig.~\ref{fig2} are clearly distinct, some qualitative similarities can also be observed. First, a strong dispersive wave like feature at around 1540~nm is apparent in each case. As highlighted by the red dashed vertical lines, the position of the feature can be accurately predicted by the well-known dispersive wave resonance condition (see~\cite{jang_observation_2014, milian_soliton_2014} and Methods). Second, each of the spectra display clear modulations that are particularly evident on the long-wavelength side. These modulations are linked to the solitons' complex, multi-peak temporal profiles, and they offer a convenient fingerprint to identify the different states. In what follows, we will refer to the different states in terms of the number of peaks in their temporal profile ($N$), deduced experimentally from the number of modulation minima on the long-wavelength edge of the spectrum ($N-1$).

\subsection{Signatures of collapsed snaking}

\noindent Theory predicts that CSs in the normal dispersion regime are organized in a bifurcation structure known as collapsed snaking (see~\cite{parra-rivas_coexistence_2017} and Supplementary Information). A defining feature of such a structure is that several distinct solitons may coexist for identical system parameters, but the range of parameters over which they exist is different. To systematically test this prediction, we have performed extensive experiments so as to map the range of existence of the different soliton states shown in Fig.~\ref{fig2}. Since it can be straightforwardly adjusted in our experiments, we use the cavity detuning as a control parameter [see Methods].

The pseudo-color plot in Fig.~\ref{fig3}(a) shows an example of the spectra measured for a soliton with three peaks in its temporal profile. 704 individually recorded spectra are shown over the entire range of the soliton's existence, visualising how the soliton spectrum broadens and the dispersive wave features shift away from the pump as the detuning increases. As the detuning is increased (decreased) beyond the values shown in Fig.~\ref{fig3}(a), the intracavity field switches to the cw state (fluctuating MI state), thus causing the soliton to cease to exist.

To plot bifurcation-like diagrams, we integrate over each spectra (excluding the pump) so as to obtain a quantity proportional to the soliton energy that can be plotted as a function of detuning. Figure~\ref{fig3}(b) shows the resulting curves for each of the soliton states shown in Fig.~\ref{fig2}. As can be seen, the different states (i) coexist for a range of detunings and (ii) exhibit different ranges of existence, with the 4- and 5-peak states existing over a narrower range than the 2- and 3-peak states. These observations are consistent with the solitons' hypothesised collapsed snaking bifurcation structure~\cite{parra-rivas_coexistence_2017}.

To compare the experimentally observed range of soliton existence with theoretical predictions, in Fig.~\ref{fig3}(c) we show the energies of the different soliton states as derived from mean-field modelling (excluding the pump as in our experiments, see also Supplementary Information). The theoretical results show very good qualitative agreement with our experimental observations, yet a number of quantitative discrepancies can be observed. We note in particular that, in our experiments, the lower boundary of soliton existence is delineated by an abrupt and spontaneous switching of the entire intracavity field to a fluctuating MI state, which is not captured by theory. Furthermore, in contrast to theory, our experiments suggest that the lower-boundary of existence of the two-peak soliton is significantly higher than the other solitons. A detailed study is beyond the scope of our present work, but we speculate that these discrepancies stem from a variety of experimental imperfections (e.g. inhomogeneities on the nanosecond driving pulses) as well as from the inability of the theoretical model to account for Kelly-like sidebands. Regardless, the experimental results presented in Fig.~\ref{fig3}(b) provide clear evidence of the collapsed snaking bifurcation structure of the bright CS structures in the normal dispersion regime.

\subsection{Single-peak CSs in the normal dispersion regime}
\noindent The results presented above pertain to bright soliton structures characterised by multiple peaks in their temporal profile [see Fig.~\ref{fig2}]. Interestingly, it is only such multi-peak structures that were identified in the theoretical study of Parra-Rivas et al~\cite{parra-rivas_coexistence_2017}, raising the natural question whether it is possible to sustain, in a Kerr resonator driven in the normal dispersion regime, single-peak bright CSs reminiscent of the conventional CSs in the anomalous dispersion regime.

The answer to the question posed above is yes. As a matter of fact, already in 2014, Mili\'an and Skryabin discovered via numerical simulations that higher-order dispersion can extend the range of existence of a conventional CS into the regime of normal dispersion driving~\cite{milian_soliton_2014}. In our experiments, we find that such single-peak, bright CSs can be readily excited when increasing the driving power to about 20~W. However, once excited, it is possible to reduce the driving power level without destroying the soliton (provided that the detuning is simultaneously reduced).

\begin{figure}[!t]
 \centering
  \includegraphics[width = \columnwidth, clip=true]{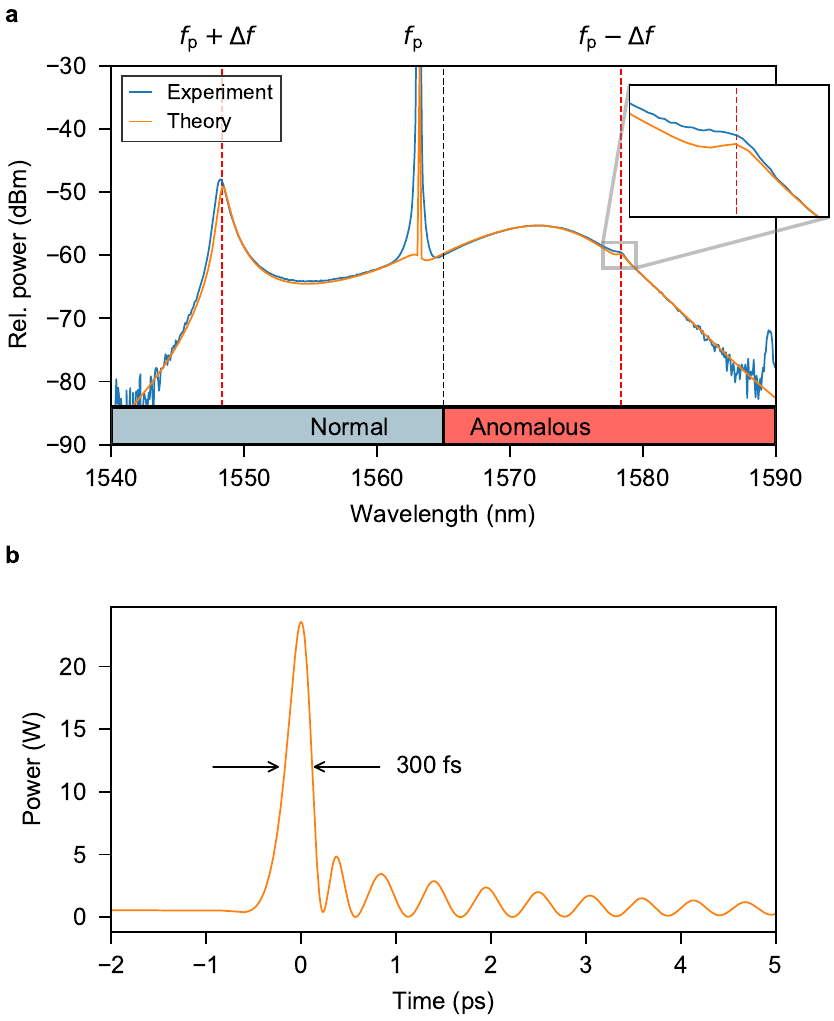}
 \caption{Experimental observation of a single-peak CS with normal-dispersion driving. \textbf{a,} Experimentally measured (solid blue curve) and theoretically predicted (solid orange curve) spectra for a driving wavelength and peak power of 1563~nm and 6.7~W, respectively. Black dashed line indicates the zero-dispersion wavelength of 1565~nm, whilst red dashed lines indicate the theoretically predicted dispersive wave positions, with $\Delta f$ the corresponding frequency shift from the pump frequency $f_\mathrm{p}$. Inset shows a zoom around the weak dispersive wave at 1578 nm. \textbf{b,} Simulated temporal profile corresponding to the theoretical spectrum shown in \textbf{a}. All parameters as in Fig.~\ref{fig3} except for the linear cavity detuning which was set to $\delta_0 = 0.8~\mathrm{rad}$.}
 \label{fig4}
 \vskip-10pt
\end{figure}

Figure~\ref{fig4}(a) shows an experimentally measured spectrum characteristic of a single-peak bright CS, obtained with the driving wavelength in the normal dispersion regime at 1563~nm (2~nm below the ZDW). Superimposed with the experimental data is the corresponding numerically simulated spectrum, whilst Fig.~\ref{fig4}(b) shows the numerically simulated temporal profile. As can be seen, the spectral profile is highly reminiscent of a conventional CS generated when driving close to the ZDW in the anomalous dispersion regime~\cite{brasch_photonic_2016, coen_modeling_2013}. The resemblance can be readily understood by noting that, although the pump resides in the normal dispersion regime, the soliton itself in fact experiences predominantly anomalous dispersion, being centred at around 1572~nm (7~nm above the ZDW). This is made possible by the emission of a strong short-wavelength dispersive wave at about 1545~nm, which causes the soliton to spectrally recoil towards the long-wavelength anomalous dispersion regime~\cite{milian_soliton_2014, coen_modeling_2013}. Accordingly, the dispersive wave and the soliton  rather remarkably correspond to a single symbiotic structure, where one cannot exist without the other. It should also be clear how the existence of such a structure is fundamentally enabled by third-order dispersion, which underpins the phase-matched energy flow between the soliton and the dispersive wave.

In addition to the strong short-wavelength dispersive wave peak at about 1545~nm, both the experimentally recorded and numerically simulated spectra show a small but noticeable long-wavelength spectral feature at about 1580~nm [see inset of Fig.~\ref{fig4}(a)]. This latter feature corresponds in fact to the symmetrically detuned (with respect to the pump frequency) counterpart of the strong dispersive wave at 1545~nm. It is a known theoretical result (\cite{jang_observation_2014, milian_soliton_2014}; see also Methods) that, in cavity systems, dispersive waves will always be generated in pairs, but until now, experimental observations have remained elusive due to the small amplitude of one the constituents of the pair. To the best of our knowledge, the results presented in Fig.~\ref{fig4}(a) correspond to the first experimental observation of a dispersive wave pair as predicted by theory.

\begin{figure}[!t]
 \centering
  \includegraphics[width = \columnwidth, clip=true]{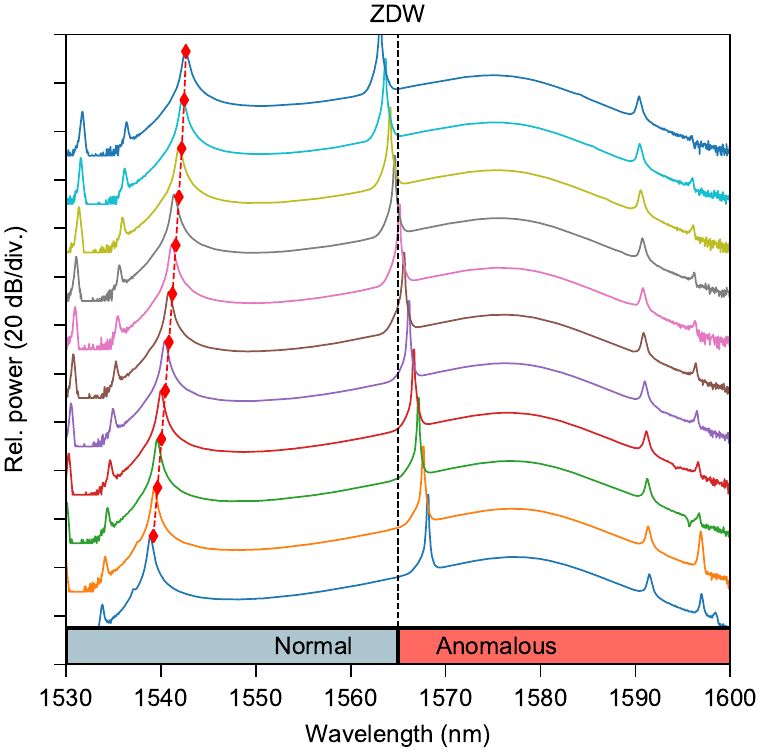}
 \caption{Experimental results showing how the single-peak soliton spectrum changes as the driving wavelength is tuned across the ZDW. The different curves show spectra measured for driving wavelengths ranging from 1563~nm (top) to 1568~nm (bottom) in 0.5~nm intervals. The dashed black line indicates the ZDW of 1565~nm, whilst the red diamonds indicate the positions of the theoretically predicted (short-wavelength) dispersive waves. All measurements use a driving pulse peak power of 15~W and a linear cavity detuning $\delta_0 = 1.5~\mathrm{rad}$.}
 \label{fig5}
 \vskip-10pt
\end{figure}

The experimental results described above corroborate the hypothesis that, by enabling dispersive wave emission and concomitant spectral recoil, third-order dispersion can extend the existence of conventional anomalous dispersion CSs into the regime of normal dispersion driving~\cite{milian_soliton_2014}. To gain more insights, we have repeated the experiment at various wavelengths across the ZDW, with Fig.~\ref{fig5} showing an assortment of the spectra recorded. As can be seen, the main consequence of tuning the pump wavelength is the shifting of the short-wavelength dispersive wave feature, which is well-predicted by theory (red diamonds). Whilst not readily visible in the plot, we also remark that, as the pump wavelength increases beyond the ZDW, the long-wavelength dispersive wave feature discussed above [see also Fig.~\ref{fig4}(a)] shifts to longer  wavelengths and becomes essentially unobservable, arguably explaining why this feature has not been observed in previous experiments. (Note that the peaks beyond 1590~nm in Fig.~\ref{fig5} correspond to Kelly-like sidebands.) We must emphasize, however, that it is the complete absence of any qualitative change in the soliton spectrum as the driving field tunes across the ZDW that is the main result to be gleaned from Fig.~\ref{fig5}. Indeed, these measurements unequivocally demonstrate that the single-peak structures manifesting themselves under conditions of normal and anomalous dispersion driving correspond to the very same CS, whose range of existence spans across the ZDW.

\subsection{Discussion}

\noindent The fact that single-peak CSs can exist under conditions of normal dispersion driving raises interesting questions on the relationship between solitons in the normal and anomalous dispersion regimes. Specifically, it is fascinating to speculate whether the bright multi-peak structures reported in Fig.~\ref{fig2} (and theoretically studied in~\cite{parra-rivas_coexistence_2017}) could be understood as bound states of individual single-peak solitons that extend from the anomalous dispersion regime. On the one hand, third-order dispersion is well-known to permit robust CS bound states~\cite{wang_universal_2017}, and we find that the single-peak structures are part of the same collapsed snaking bifurcation structure as the multi-peak structures [see Supplementary Information]. But on the other hand, the characteristically close proximity of the peaks as seen in Fig.~\ref{fig2} speaks against the presence of individual (albeit bound) structures, as does the fact that the multi-peak structures can exist over a wider range of parameters than their single-peak counterparts [see Supplementary Information]. Further study is required to fully understand where and how the different soliton states (and their physical origins) meet.

To the best of our knowledge, the results obtained in our work represent the first direct experimental observations of bright CSs enabled by third-order dispersion, and they provide clear signatures of the solitons' collapsed snaking bifurcation structure. Moreover, our findings demonstrate that higher-order dispersion can extend the range of existence of conventional single-peak bright CSs into the regime of normal dispersion driving, as well as give rise to symmetric dispersive wave pairs. Taken together, our work confirms a string of theoretical predictions~\cite{parra-rivas_coexistence_2017, milian_soliton_2014} that have remained unobserved in direct experiments until now.

Besides the verification of earlier theoretical predictions, another major outcome of our work is the demonstration of a robust experimental platform that allows for the systematic study of novel Kerr cavity dynamics that manifest themselves under the influence of higher-order dispersion. In this context, our experiments have revealed a rich diversity of phenomena that have hitherto not been studied neither in experiment nor theory. These include discontinuities in the soliton existence range at high pump powers, as well as persisting coexistence of incoherent modulation instability and coherent soliton states. Future works will be dedicated for the comprehensive study of such dynamics.

Before closing, we note that the solitons whose spectra are shown in Fig.~\ref{fig5} are modelled to have a temporal duration as short as 220~fs (full-width at half-maximum), which is the shortest duration ever achieved for CSs in a passive fibre ring resonator. We envisage that the duration can be further reduced through judicious optimisation of the cavity length and losses. When combined with demonstrated techniques to manipulate CSs in macroscopic fibre ring resonators~\cite{jang_temporal_2015,jang_all-optical_2016}, our work could pave the way for a novel source of ultrashort pulses with widely tunable repetition rate that can be easily locked to an external RF reference. Finally, by showing that bright CSs can exist (and be spontaneously excited) under conditions of normal dispersion driving, our work may provide new degrees of freedom for the design of dispersion-engineered microresonator frequency combs.

\section*{Methods}

\small

\subparagraph*{\hskip-10pt Additional resonator characteristics.} The fibre ring resonator used in our experiments is made entirely of dispersion-shifted fibre with measured zero-dispersion wavelength of 1565~nm [see Supplementary Information for details regarding dispersion measurement]. The resonator incorporates two couplers that are custom-made from the same DSF as the rest of the cavity so as to minimise dispersion fluctuations. We estimate the resonator to have a Kerr nonlinearity coefficient of $\gamma = 1.5~\mathrm{W}^{-1}\mathrm{km}^{-1}$. To measure the cavity finesse, we used a narrow-linewidth distributed feedback fibre laser (NKT Adjustik) to drive the cavity, and scanned the laser frequency over several free-spectral ranges. The finesse can then be readily obtained from the measured cavity output signal. Note that, because of its finite linewidth, the external cavity diode laser (ECDL) used in our main experiments cannot be used to reliably measure the finesse by scanning over several free-spectral ranges. Rather, as mentioned in the main text, the ECDL can only be used to coherently drive the resonator when the detuning is locked using the Pound-Drever-Hall scheme.

\subparagraph*{\hskip-10pt Measurement of laser linewidth.} We measured the linewidth of the ECDL using a homodyne detection scheme. Here the laser beam is first divided into two parts, with one part frequency-shifted using an acousto-optic modulator and the other part passed through 30~km of fibre so as to de-correlate the two paths. Superimposing the beams on a photodetector allows the linewidth to be estimated from the recorded beat signal. We find that, when not actively locked to the fibre ring resonator, the ECDL has a free-running linewidth of about 80~kHz. In contrast, when the laser is locked to the resonator using the PDH scheme, the linewidth reduces to about 60~kHz, which is close to the detection limit with a 30~km delay stage. Moreover, we must note that the measurement is of the absolute linewidth and therefore includes contributions from the fluctuations of the cavity resonances. The fact that we are able to observe coherent cavity dynamics over a wide range of parameters demonstrates that the linear detuning is well stabilized, and the laser linewidth sufficiently narrowed with respect to the cavity resonance.

\subparagraph*{\hskip-10pt Dispersion coefficients.} As mentioned in our main text, we measured the resonator to have a zero-dispersion wavelength of $\lambda_\mathrm{ZDW} = 1565.4~\mathrm{nm}$ and third- and fourth-order dispersion coefficients of $\beta_{3,\text{ZDW}} = 0.135~\mathrm{ps^3/km}$ and \mbox{$\beta_{4,\text{ZDW}} = -9\times10^{-4}~\mathrm{ps^4/km}$} at the ZDW, respectively [see also Supplementary Information]. These coefficients readily allow us to evaluate the dispersion coefficients at all other wavelengths $\lambda$ using the following relations:
\begin{align*}
\beta_2 &= \beta_{3,\text{ZDW}}\Omega + \frac{\beta_{4,\text{ZDW}}}{2}\Omega^2, \\
\beta_3 &= \beta_{3,\text{ZDW}} + \beta_{4,\text{ZDW}}\Omega, \\
\beta_4 &= \beta_{4,\text{ZDW}},
\end{align*}
where $\Omega = 2\pi c\left[\lambda^{-1} - \lambda_\mathrm{ZDW}^{-1} \right]$ with $c$ the speed of light in vacuum.

The relative strength of third-order dispersion, which underpins the bright solitons studied in our work, can be estimated using the dimensionless quantity~\cite{jang_observation_2014, parra-rivas_coexistence_2017}:
\begin{equation}
d_3 = \sqrt{\frac{2\pi}{L\mathcal{F}}}\left(\frac{\beta_3}{(3|\beta_2|)^{3/2}}\right),
\end{equation}
where $L$ and $\mathcal{F}$ are the resonator length and finesse, respectively. By allowing us to operate arbitrarily close to the ZDW, and hence control the value of $|\beta_2|$, our platform gives access to a wide range of $d_3$ coefficients. When driving at 1563~nm [see Figs.~\ref{fig2}--\ref{fig4}], we estimate $d_3\approx 0.9$; when scanning the pump wavelength across the ZDW to obtain data shown in Fig.~\ref{fig5}, we estimate $d_3$ to vary from 0.7 to 67.

\subparagraph*{\hskip-10pt Measurement of bifurcation curves.} To experimentally measure how the bright solitons change as a function of the detuning, we first lock the cavity detuning at an appropriate value, and perturb the cavity until a given soliton is excited. A computer-automated algorithm then adiabatically increases the detuning by changing the frequency applied on the acousto-optic modulator in small discrete increments, recording the soliton spectrum at each step. Once the upper limit of the soliton existence is deduced in this manner, we repeat the procedure but now incrementally reducing the detuning until the lower limit of existence is identified. We must emphasize that the detuning is kept actively stabilized through the entire measurement. These discrete detuning scans are repeated several times for each soliton so as to ensure that their full range of existence is captured.

\subparagraph*{\hskip-10pt Theoretical model.} The theoretical results presented in our manuscript were obtained using the well-known generalized mean-field equation of Kerr cavities~\cite{coen_modeling_2013, haelterman_dissipative_1992, wang_stimulated_2018}. For completeness, our model includes dispersion up to fourth order as well as stimulated Raman scattering [see Supplementary Information for full details], yet we note that a simpler model that limits dispersion to third order and ignores Raman effects is sufficient to describe the principal physics of the bright solitons studied in our work.  We find the steady-state solutions of our theoretical model using a multidimensional Newton-Raphson root-finding algorithm, and apply numerical continuation to study the solutions' bifurcations. All the theoretical results shown use the corresponding experimental parameters.

\subparagraph*{\hskip-10pt Dispersive wave positions.} The spectral positions of dispersive waves emitted by Kerr cavity solitons (CSs) can be found from the roots of the following characteristic equation~\cite{jang_observation_2014, milian_soliton_2014}:
	\begin{equation}
	\small
	\begin{split}
-\frac{\pi}{\mathcal{F}} &+i\frac{\beta_3 L}{3!}Q^3 - iVQ \\
&\pm i \sqrt{\left(2\gamma L P_0 -\delta_0 +\frac{\beta_2 L}{2!}Q^2 +\frac{\beta_4 L}{4!}Q^4 \right)^2 - (\gamma L P_0)^2} = 0.
	\end{split}
	\label{dwpos}
	\end{equation}
Here $V$ represents the group-delay accumulated by the CSs with respect to the driving field over one round trip, $P_0$ is the power level of the (quasi) continuous wave background on top of which the solitons sit, $\delta_0$ is the phase detuning, and $Q$ is a complex frequency whose real part yields the dispersive wave frequency shift from the pump, $\Delta f = \text{Re}[Q]/(2\pi)$.

The two signs in front of the square root in Eq.~\eqref{dwpos} signal that, in a cavity geometry, dispersive waves come in pairs. It is straightforward to show that, if $Q_1$ is a solution for one of the signs, then the solution for the second sign is given by $Q_2 = -Q_1^*$. This implies that the corresponding frequency shifts $\Delta f_1 = -\Delta f_2$, thus demonstrating that the two dispersive waves are symmetrically detuned with respect to the pump.

For parameters pertinent to our study, the second term inside the square root of Eq.~\eqref{dwpos} is small. Focussing on the $+$ sign in front to of the square root, we may approximate the equation as a polynomial in $Q$~\cite{jang_observation_2014}:
\begin{equation}
\frac{\beta_4 L}{4!} Q^4 + \frac{\beta_3 L}{3!} Q^3 + \frac{\beta_2 L}{2!} Q^2 - VQ + \left[(2\gamma L P_0 -\delta_0) +i\frac{\pi}{\mathcal{F}}\right] = 0.
\end{equation}
All the theoretically predicted dispersive wave frequency shifts shown in our work were obtained by finding the roots of this polynomial, with the background power level $P_0$ obtained from the usual cubic polynomial characteristic of dispersive bistability, and the CS group-delay $V$ extracted from numerical simulations that use experimental parameters.

\section*{Acknowledgements}
\noindent We acknowledge financial support from the Marsden Fund, the Rutherford Discovery Fellowships, and the James Cook Fellowships of the Royal Society of New Zealand. We also acknowledge useful discussions and help from Alexander Nielsen and Ian Hendry.



\begin{thebibliography}{100}
\newcommand{\enquote}[1]{``#1''}

\bibitem{laughlin_solitary_1983}
D.~W.~M. Laughlin, J.~V. Moloney, and A.~C. Newell, \enquote{Solitary {Waves}
  as {Fixed} {Points} of {Infinite}-{Dimensional} {Maps} in an {Optical}
  {Bistable} {Ring} {Cavity},} Phys. Rev. Lett. \textbf{51}, 75--78 (1983).

\bibitem{firth_optical_1996}
W.~J. Firth and A.~J. Scroggie, \enquote{Optical {Bullet} {Holes}: {Robust}
  {Controllable} {Localized} {States} of a {Nonlinear} {Cavity},} Phys. Rev.
  Lett. \textbf{76}, 1623--1626 (1996).

\bibitem{spinelli_spatial_1998}
L.~Spinelli, G.~Tissoni, M.~Brambilla, F.~Prati, and L.~A. Lugiato,
  \enquote{Spatial solitons in semiconductor microcavities,} Phys. Rev. A
  \textbf{58}, 2542--2559 (1998).

\bibitem{firth_cavity_2002}
W.~J. Firth and C.~O. Weiss, \enquote{Cavity and {Feedback} {Solitons},} Opt.
  Photon. News \textbf{13}, 54--58 (2002).

\bibitem{ackemann_chapter_2009}
T.~Ackemann, W.~Firth, and G.-L. Oppo, \enquote{Chapter 6 {Fundamentals} and
  {Applications} of {Spatial} {Dissipative} {Solitons} in {Photonic}
  {Devices},} in P.~R. B. a. C. C.~L. E.~Arimondo, editor, \enquote{Advances
  {In} {Atomic}, {Molecular}, and {Optical} {Physics},} volume Volume 57, pages
  323--421, Academic Press (2009).

\bibitem{barland_cavity_2002}
S.~Barland, J.~R. Tredicce, M.~Brambilla, L.~A. Lugiato, S.~Balle, M.~Giudici,
  T.~Maggipinto, L.~Spinelli, G.~Tissoni, T.~Kn{\"o}dl, M.~Miller, and
  R.~J{\"a}ger, \enquote{Cavity solitons as pixels in semiconductor
  microcavities,} Nature \textbf{419}, 699--702 (2002).

\bibitem{wabnitz_suppression_1993}
S.~Wabnitz, \enquote{Suppression of interactions in a phase-locked soliton optical memory,}
  Opt. Lett. \textbf{18}, 601--603 (1993).

\bibitem{leo_temporal_2010}
F.~Leo, S.~Coen, P.~Kockaert, S.-P. Gorza, P.~Emplit, and M.~Haelterman,
  \enquote{Temporal cavity solitons in one-dimensional {Kerr} media as bits in
  an all-optical buffer,} Nat. Photon. \textbf{4}, 471--476 (2010).

\bibitem{leo_dynamics_2013}
F.~Leo, L.~Gelens, P.~Emplit, M.~Haelterman, and S.~Coen, \enquote{Dynamics of
  one-dimensional {Kerr} cavity solitons,} Opt. Express \textbf{21}, 9180--9191
  (2013).

\bibitem{jang_observation_2014}
J.~K. Jang, M.~Erkintalo, S.~G. Murdoch, and S.~Coen, \enquote{Observation of
  dispersive wave emission by temporal cavity solitons,} Opt. Lett.
  \textbf{39}, 5503--5506 (2014).

\bibitem{jang_temporal_2015}
J.~K. Jang, M.~Erkintalo, S.~Coen, and S.~G. Murdoch, \enquote{Temporal
  tweezing of light through the trapping and manipulation of temporal cavity
  solitons,} Nat. Commun. \textbf{6} (2015).

\bibitem{jang_all-optical_2016}
J.~K. Jang, M.~Erkintalo, J.~Schr{\"o}der, B.~J. Eggleton, S.~G. Murdoch, and
  S.~Coen, \enquote{All-optical buffer based on temporal cavity solitons
  operating at 10 {Gb}/s,} Opt. Lett. \textbf{41}, 4526 (2016).

\bibitem{anderson_coexistence_2017}
M.~Anderson, Y.~Wang, F.~Leo, S.~Coen, M.~Erkintalo, and S.~G. Murdoch,
  \enquote{Coexistence of {Multiple} {Nonlinear} {States} in a {Tristable}
  {Passive} {Kerr} {Resonator},} Phys. Rev. X \textbf{7}, 031031 (2017).

\bibitem{nielsen_coexistence_2019}
A.~U. Nielsen, B.~Garbin, S.~Coen, S.~G. Murdoch, and M.~Erkintalo,
  \enquote{Coexistence and {Interactions} between {Nonlinear} {States} with
  {Different} {Polarizations} in a {Monochromatically} {Driven} {Passive}
  {Kerr} {Resonator},} Phys. Rev. Lett. \textbf{123}, 013902 (2019).

\bibitem{jang_ultraweak_2013}
J.~K. Jang, M.~Erkintalo, S.~G. Murdoch, and S.~Coen, \enquote{Ultraweak
  long-range interactions of solitons observed over astronomical distances,}
  Nat. Photon. \textbf{7}, 657--663 (2013).

\bibitem{anderson_observations_2016}
M.~Anderson, F.~Leo, S.~Coen, M.~Erkintalo, and S.~G. Murdoch,
  \enquote{Observations of spatiotemporal instabilities of temporal cavity
  solitons,} Optica \textbf{3}, 1071 (2016).

\bibitem{xue_super-efficient_2019}
X.~Xue, X.~Zheng, and B.~Zhou, \enquote{Super-efficient temporal solitons in
  mutually coupled optical cavities,} Nat. Photon. \textbf{13}, 616--622
  (2019).

\bibitem{herr_temporal_2014}
T.~Herr, V.~Brasch, J.~D. Jost, C.~Y. Wang, N.~M. Kondratiev, M.~L. Gorodetsky,
  and T.~J. Kippenberg, \enquote{Temporal solitons in optical microresonators,}
  Nat. Photon. \textbf{8}, 145--152 (2014).

\bibitem{delhaye_optical_2007}
P.~Del'Haye, A.~Schliesser, O.~Arcizet, T.~Wilken, R.~Holzwarth, and T.~J.
  Kippenberg, \enquote{Optical frequency comb generation from a monolithic
  microresonator,} Nature \textbf{450}, 1214--1217 (2007).

\bibitem{brasch_photonic_2016}
V.~Brasch, M.~Geiselmann, T.~Herr, G.~Lihachev, M.~H.~P. Pfeiffer, M.~L.
  Gorodetsky, and T.~J. Kippenberg, \enquote{Photonic chip{\textendash}based
  optical frequency comb using soliton {Cherenkov} radiation,} Science
  \textbf{351}, 357--360 (2016).

\bibitem{obrzud_temporal_2017}
E.~Obrzud, S.~Lecomte, and T.~Herr, \enquote{Temporal solitons in
  microresonators driven by optical pulses,} Nat. Photon. \textbf{11}, 600
  (2017).

\bibitem{cole_kerr-microresonator_2018}
D.~C. Cole, J.~R. Stone, M.~Erkintalo, K.~Y. Yang, X.~Yi, K.~J. Vahala, and
  S.~B. Papp, \enquote{Kerr-microresonator solitons from a chirped background,}
  Optica \textbf{5}, 1304--1310 (2018).

\bibitem{lucas_spatial_2018}
E.~Lucas, G.~Lihachev, R.~Bouchand, N.~G. Pavlov, A.~S. Raja, M.~Karpov, M.~L.
  Gorodetsky, and T.~J. Kippenberg, \enquote{Spatial multiplexing of soliton
  microcombs,} Nat. Photon. \textbf{12}, 699--705 (2018).

\bibitem{jang_synchronization_2018}
J.~K. Jang, A.~Klenner, X.~Ji, Y.~Okawachi, M.~Lipson, and A.~L. Gaeta,
  \enquote{Synchronization of coupled optical microresonators,} Nat.
  Photon. \textbf{12}, 688--693 (2018).

\bibitem{kippenberg_dissipative_2018}
T.~J. Kippenberg, A.~L. Gaeta, M.~Lipson, and M.~L. Gorodetsky,
  \enquote{Dissipative {Kerr} solitons in optical microresonators,} Science
  \textbf{361}, eaan8083 (2018).

\bibitem{marin-palomo_microresonator-based_2017}
P.~Marin-Palomo, J.~N. Kemal, M.~Karpov, A.~Kordts, J.~Pfeifle, M.~H.~P.
  Pfeiffer, P.~Trocha, S.~Wolf, V.~Brasch, M.~H. Anderson, R.~Rosenberger,
  K.~Vijayan, W.~Freude, T.~J. Kippenberg, and C.~Koos,
  \enquote{Microresonator-based solitons for massively parallel coherent
  optical communications,} Nature \textbf{546}, 274 (2017).

\bibitem{trocha_ultrafast_2018}
P.~Trocha, M.~Karpov, D.~Ganin, M.~H.~P. Pfeiffer, A.~Kordts, S.~Wolf,
  J.~Krockenberger, P.~Marin-Palomo, C.~Weimann, S.~Randel, W.~Freude, T.~J.
  Kippenberg, and C.~Koos, \enquote{Ultrafast optical ranging using
  microresonator soliton frequency combs,} Science \textbf{359}, 887--891
  (2018).

\bibitem{suh_soliton_2018}
M.-G. Suh and K.~J. Vahala, \enquote{Soliton microcomb range measurement,}
  Science \textbf{359}, 884--887 (2018).

\bibitem{suh_microresonator_2016}
M.-G. Suh, Q.-F. Yang, K.~Y. Yang, X.~Yi, and K.~J. Vahala,
  \enquote{Microresonator soliton dual-comb spectroscopy,} Science
  \textbf{354}, 600--603 (2016).

\bibitem{dutt_-chip_2018}
A.~Dutt, C.~Joshi, X.~Ji, J.~Cardenas, Y.~Okawachi, K.~Luke, A.~L. Gaeta, and
  M.~Lipson, \enquote{On-chip dual-comb source for spectroscopy,} Sci.
  Adv. \textbf{4}, e1701858 (2018).

\bibitem{spencer_optical-frequency_2018}
D.~T. Spencer, T.~Drake, T.~C. Briles, J.~Stone, L.~C. Sinclair, C.~Fredrick,
  Q.~Li, D.~Westly, B.~R. Ilic, A.~Bluestone, N.~Volet, T.~Komljenovic,
  L.~Chang, S.~H. Lee, D.~Y. Oh, M.-G. Suh, K.~Y. Yang, M.~H.~P. Pfeiffer,
  T.~J. Kippenberg, E.~Norberg, L.~Theogarajan, K.~Vahala, N.~R. Newbury,
  K.~Srinivasan, J.~E. Bowers, S.~A. Diddams, and S.~B. Papp, \enquote{An
  optical-frequency synthesizer using integrated photonics,} Nature
  \textbf{557}, 81--85 (2018).

\bibitem{carmon_dynamical_2004}
T.~Carmon, L.~Yang, and K.~J. Vahala, \enquote{Dynamical thermal behavior and
  thermal self-stability of microcavities,} Opt. Express \textbf{12},
  4742--4750 (2004).

\bibitem{herr_mode_2014}
T.~Herr, V.~Brasch, J.~D. Jost, I.~Mirgorodskiy, G.~Lihachev, M.~L. Gorodetsky,
  and T.~J. Kippenberg, \enquote{Mode {Spectrum} and {Temporal} {Soliton}
  {Formation} in {Optical} {Microresonators},} Phys. Rev. Lett. \textbf{113},
  123901 (2014).

\bibitem{cole_soliton_2017}
D.~C. Cole, E.~S. Lamb, P.~Del{\textquoteright}Haye, S.~A. Diddams, and S.~B.
  Papp, \enquote{Soliton crystals in {Kerr} resonators,} Nat. Photon.
  \textbf{11}, 671--676 (2017).

\bibitem{karpov_dynamics_2019}
M.~Karpov, M.~H.~P. Pfeiffer, H.~Guo, W.~Weng, J.~Liu, and T.~J. Kippenberg,
  \enquote{Dynamics of soliton crystals in optical microresonators,} Nat.
  Phys. \textbf{15}, 1071--1077 (2019).

\bibitem{guo_universal_2016}
H.~Guo, M.~Karpov, E.~Lucas, A.~Kordts, M.~H.~P. Pfeiffer, V.~Brasch,
  G.~Lihachev, V.~E. Lobanov, M.~L. Gorodetsky, and T.~J. Kippenberg,
  \enquote{Universal dynamics and deterministic switching of dissipative {Kerr}
  solitons in optical microresonators,} Nat. Phys. \textbf{13}, 94--102 (2017).

\bibitem{yi_single-mode_2017}
X.~Yi, Q.-F. Yang, X.~Zhang, K.~Y. Yang, X.~Li, and K.~Vahala,
  \enquote{Single-mode dispersive waves and soliton microcomb dynamics,} Nat.
  Commun. \textbf{8}, 14869 (2017).

\bibitem{xue_mode-locked_2015}
X.~Xue, Y.~Xuan, Y.~Liu, P.-H. Wang, S.~Chen, J.~Wang, D.~E. Leaird, M.~Qi, and
  A.~M. Weiner, \enquote{Mode-locked dark pulse {Kerr} combs in
  normal-dispersion microresonators,} Nat. Photon. \textbf{9}, 594--600 (2015).

\bibitem{parra-rivas_origin_2016}
P.~Parra-Rivas, D.~Gomila, E.~Knobloch, S.~Coen, and L.~Gelens, \enquote{Origin
  and stability of dark pulse {Kerr} combs in normal dispersion resonators,}
  Opt. Lett. \textbf{41}, 2402--2405 (2016).

\bibitem{xue_microresonator_2017}
X.~Xue, P.-H. Wang, Y.~Xuan, M.~Qi, and A.~M. Weiner, \enquote{Microresonator
  {Kerr} frequency combs with high conversion efficiency,} Laser Photonics Rev. \textbf{11}, 1600276 (2017).

\bibitem{fulop_high-order_2018}
A.~F{\"u}l{\"o}p, M.~Mazur, A.~Lorences-Riesgo, {\'O}.~B. Helgason, P.-H. Wang,
  Y.~Xuan, D.~E. Leaird, M.~Qi, P.~A. Andrekson, A.~M. Weiner, and
  V.~Torres-Company, \enquote{High-order coherent communications using
  mode-locked dark-pulse {Kerr} combs from microresonators,} Nat.
  Commun. \textbf{9}, 1--8 (2018).

\bibitem{helgason_superchannel_2019}
{\'O}.~B. Helgason, A.~F{\"u}l{\"o}p, J.~Schr{\"o}der, P.~A. Andrekson, A.~M.
  Weiner, and V.~Torres-Company, \enquote{Superchannel engineering of
  microcombs for optical communications,} J. Opt. Soc. Am. B \textbf{36}, 2013--2022 (2019).

\bibitem{lobanov_frequency_2015}
V.~E. Lobanov, G.~Lihachev, T.~J. Kippenberg, and M.~L. Gorodetsky,
  \enquote{Frequency combs and platicons in optical microresonators with normal
  {GVD},} Opt. Express \textbf{23}, 7713--7721 (2015).

\bibitem{lobanov_generation_2015}
V.~E. Lobanov, G.~Lihachev, and M.~L. Gorodetsky, \enquote{Generation of
  platicons and frequency combs in optical microresonators with normal {GVD} by
  modulated pump,} EPL \textbf{112}, 54008 (2015).

\bibitem{jang_dynamics_2016}
J.~K. Jang, Y.~Okawachi, M.~Yu, K.~Luke, X.~Ji, M.~Lipson, and A.~L. Gaeta,
  \enquote{Dynamics of mode-coupling-induced microresonator frequency combs in
  normal dispersion,} Opt. Express \textbf{24}, 28794--28803 (2016).

\bibitem{lobanov_dynamics_2017}
V.~E. Lobanov, A.~V. Cherenkov, A.~E. Shitikov, I.~A. Bilenko, and M.~L.
  Gorodetsky, \enquote{Dynamics of platicons due to third-order dispersion,}
  Eur. Phys. J. D \textbf{71}, 185 (2017).

\bibitem{huang_mode-locked_2015}
S.-W. Huang, H.~Zhou, J.~Yang, J.~F. McMillan, A.~Matsko, M.~Yu, D.-L. Kwong,
  L.~Maleki, and C.~W. Wong, \enquote{Mode-{Locked} {Ultrashort} {Pulse}
  {Generation} from {On}-{Chip} {Normal} {Dispersion} {Microresonators},} Phys.
  Rev. Lett. \textbf{114}, 053901 (2015).

\bibitem{spiess_chirped_2019}
C.~Spiess, Q.~Yang, X.~Dong, V.~G. Bucklew, and W.~H. Renninger,
  \enquote{Chirped temporal solitons in driven optical resonators,}
  arXiv:1906.12127 [physics]  (2019).

\bibitem{parra-rivas_coexistence_2017}
P.~Parra-Rivas, D.~Gomila, and L.~Gelens, \enquote{Coexistence of stable dark-
  and bright-soliton {Kerr} combs in normal-dispersion resonators,} Phys. Rev.
  A \textbf{95}, 053863 (2017).

\bibitem{copie_competing_2016}
F.~Copie, M.~Conforti, A.~Kudlinski, A.~Mussot, and S.~Trillo,
  \enquote{Competing {Turing} and {Faraday} {Instabilities} in {Longitudinally}
  {Modulated} {Passive} {Resonators},} Phys. Rev. Lett. \textbf{116}, 143901
  (2016).

\bibitem{copie_dynamics_2017}
F.~Copie, M.~Conforti, A.~Kudlinski, S.~Trillo, and A.~Mussot,
  \enquote{Dynamics of {Turing} and {Faraday} instabilities in a longitudinally
  modulated fiber-ring cavity,} Opt. Lett. \textbf{42}, 435--438 (2017).

\bibitem{wang_universal_2017}
Y.~Wang, F.~Leo, J.~Fatome, M.~Erkintalo, S.~G. Murdoch, and S.~Coen,
  \enquote{Universal mechanism for the binding of temporal cavity solitons,}
  Optica \textbf{4}, 855--863 (2017).

\bibitem{nielsen_invited_2018}
A.~U. Nielsen, B.~Garbin, S.~Coen, S.~G. Murdoch, and M.~Erkintalo,
  \enquote{Invited {Article}: {Emission} of intense resonant radiation by
  dispersion-managed {Kerr} cavity solitons,} APL Photonics \textbf{3}, 120804
  (2018).

\bibitem{milian_soliton_2014}
C.~Mili{\'a}n and D.~Skryabin, \enquote{Soliton families and resonant radiation
  in a micro-ring resonator near zero group-velocity dispersion,} Opt.
  Express \textbf{22}, 3732 (2014).

\bibitem{bessin_modulation_2017}
F.~Bessin, F.~Copie, M.~Conforti, A.~Kudlinski, and A.~Mussot,
  \enquote{Modulation instability in the weak normal dispersion region of
  passive fiber ring cavities,} Opt. Lett. \textbf{42}, 3730--3733 (2017).

\bibitem{sayson_octave-spanning_2019}
N.~L.~B. Sayson, T.~Bi, V.~Ng, H.~Pham, L.~S. Trainor, H.~G.~L. Schwefel,
  S.~Coen, M.~Erkintalo, and S.~G. Murdoch, \enquote{Octave-spanning tunable
  parametric oscillation in crystalline {Kerr} microresonators,} Nat. Photon.
  \textbf{13}, 701--706 (2019).

\bibitem{coen_modeling_2013}
S.~Coen, H.~G. Randle, T.~Sylvestre, and M.~Erkintalo, \enquote{Modeling of
  octave-spanning {Kerr} frequency combs using a generalized mean-field
  {Lugiato}{\textendash}{Lefever} model,} Opt. Lett. \textbf{38}, 37--39
  (2013).

\bibitem{haelterman_dissipative_1992}
M.~Haelterman, S.~Trillo, and S.~Wabnitz, \enquote{Dissipative modulation
  instability in a nonlinear dispersive ring cavity,} Opt. Commun. \textbf{91},
  401--407 (1992).

\bibitem{wang_stimulated_2018}
Y.~Wang, M.~Anderson, S.~Coen, S.~G. Murdoch, and M.~Erkintalo,
  \enquote{Stimulated {Raman} {Scattering} {Imposes} {Fundamental} {Limits} to
  the {Duration} and {Bandwidth} of {Temporal} {Cavity} {Solitons},} Phys. Rev.
  Lett. \textbf{120}, 053902 (2018).

\end{thebibliography}
\end{document}


\title{Supplementary information -- Experimental observations of bright dissipative Kerr cavity solitons and their collapsed snaking in a driven resonator with normal dispersion}

\author{Zongda Li$^{1,2}$}
\author{St\'ephane Coen$^{1,2}$}
\author{Stuart G. Murdoch$^{1,2}$}
\author{Miro Erkintalo$^{1,2,}$}
\email{m.erkintalo@auckland.ac.nz}
\affiliation{$^1$Department of Physics, University of Auckland, Auckland 1010, New Zealand}
\affiliation{$^2$The Dodd-Walls Centre for Photonic and Quantum Technologies, New Zealand}
	
	\begin{abstract}
		 This article contains supplementary information to the manuscript entitled ``\emph{Experimental observations of bright dissipative Kerr cavity solitons and their collapsed snaking in a driven resonator with normal dispersion}''. We present the theoretical model used to obtain the simulation results presented in the main manuscript, and provide additional theoretical and experimental details pertinent to our study.
	\end{abstract}
	
	\maketitle
\renewcommand{\theequation}{S\arabic{equation}}
	\renewcommand{\thefigure}{S\arabic{figure}}
%
%
%

\section*{Theoretical model}
\noindent The dynamics of dispersive, Kerr nonlinear ring resonators can be described in the high-finesse limit by a mean-field equation that describes the evolution of the slowly-varying electric field envelope inside the resonator over consecutive round trips~\cite{haelterman_dissipative_1992}. The numerical simulations presented in our work were obtained from the following generalized mean-field equation that takes into account high-order dispersion and stimulated Raman scattering~\cite{coen_modeling_2013,milian_solitons_2015,wang_stimulated_2018, hendry_impact_2019}:
	\begin{equation}
		\small
		\begin{split}
		t_R\frac{\partial E(t,\tau)}{\partial t}&=\left[ -\alpha-i\delta_\mathrm{0}+iL\sum_{k\geq2}\frac{\beta_k}{k!}\left(i\frac{\partial}{\partial \tau}\right)^k\right]E+\sqrt{\theta}E_{\text{in}}\\&+i\gamma L\left[(1-f_R)|E|^2+f_Rh_R(\tau)*|E|^2\right]E.
		\end{split}
		\label{LEE1}
	\end{equation}	
Here, $t$ is a slow time variable that describes the evolution of the intracavity field envelope $E(t,\tau)$ [with units of $\mathrm{W^{1/2}}$] over consecutive round trips, $\tau$ is a corresponding fast time variable that describes the envelope's temporal profile, $t_\mathrm{R}$ is the cavity round trip time, $\alpha$ corresponds to half the total power lost per round trip, $\delta_\mathrm{0}$ is the phase detuning of the driving field $E_\mathrm{in}$ from the closest cavity resonance, $L$ is the cavity round trip length, $\beta_k$ and $\gamma$ are the dispersion and Kerr nonlinearity coefficients at the pump frequency, respectively, and $\theta$ is the coupling power transmission coefficient. Finally, $h_\mathrm{R}(\tau)$ is the time-domain response function that characterizes the Raman nonlinearity of the resonator~\cite{Stolen2}, with $f_\mathrm{R}$ the corresponding Raman fraction.

The simulations presented in our main manuscript assume a Raman fraction of $f_\mathrm{R} = 0.18$ and a Raman response function $h_\mathrm{R}(\tau)$ obtained from the well-known multiple-vibrational-mode model~\cite{Hollenbeck_multiple-vibrational-mode_2002}. Dispersion is included to fourth-order, with $\beta_{3,\text{ZDW}} = 0.135~\mathrm{ps^3/km}$ and \mbox{$\beta_{4,\text{ZDW}} = -9\times10^{-4}~\mathrm{ps^4/km}$} at the zero-dispersion wavelength of $\lambda_\mathrm{ZDW} = 1565.4~\mathrm{nm}$ corresponding to experimentally measured values (see discussion below). The nonlinearity coefficient we estimate to be \mbox{$\gamma=1.5~\mathrm{W^{-1}km^{-1}}$}, the resonator length $L = 26~\mathrm{m}$, input coupling coefficient $\theta = 0.05$ and cavity losses $\alpha = 0.07$ (corresponding to the experimentally measured finesse of 45). Our experiments consider different values of cavity detunings $\delta_0$ and pump power $P_\mathrm{in} = |E_\mathrm{in}|^2$; these are quoted in our main manuscript and used in the corresponding simulations.

Before proceeding, we must emphasize that our numerical model includes fourth-order dispersion and stimulated Raman scattering solely for the sake of completeness. The existence of the bright solitons studied in our work, as well as their salient characteristics, can be well described even without the inclusion of these effects. As discussed below, the main impact of fourth-order dispersion is to facilitate the excitation of bright structures. The main impact of Raman scattering appears to be the extension of the range of soliton existence; detailed investigation of this phenomenon is left for future work.


\section*{Illustrative simulation results and Collapsed snaking}

\noindent As described in detail in ref.~\cite{parra-rivas_coexistence_2017}, both dark and bright localized structures in resonators with normal dispersion can be explained as interlocked switching waves (also called fronts or domain walls) that connect the lower and upper homogeneous states of the bistable cavity system. These localized structures are arranged in a bifurcation structure referred to as collapsed snaking, which is characterized by the partial coexistence of several structures with distinct widths. Here we present illustrative results pertaining to solitons and their collapsed snaking under conditions of normal dispersion driving ($\beta_2>0$) and in the presence of third-order dispersion ($\beta_3\neq0$). For the sake of simplicity, we ignore fourth-order dispersion and stimulated Raman scattering ($\beta_4 = f_\mathrm{R} = 0$)

\begin{figure*}[t!]
	\includegraphics[width=\linewidth]{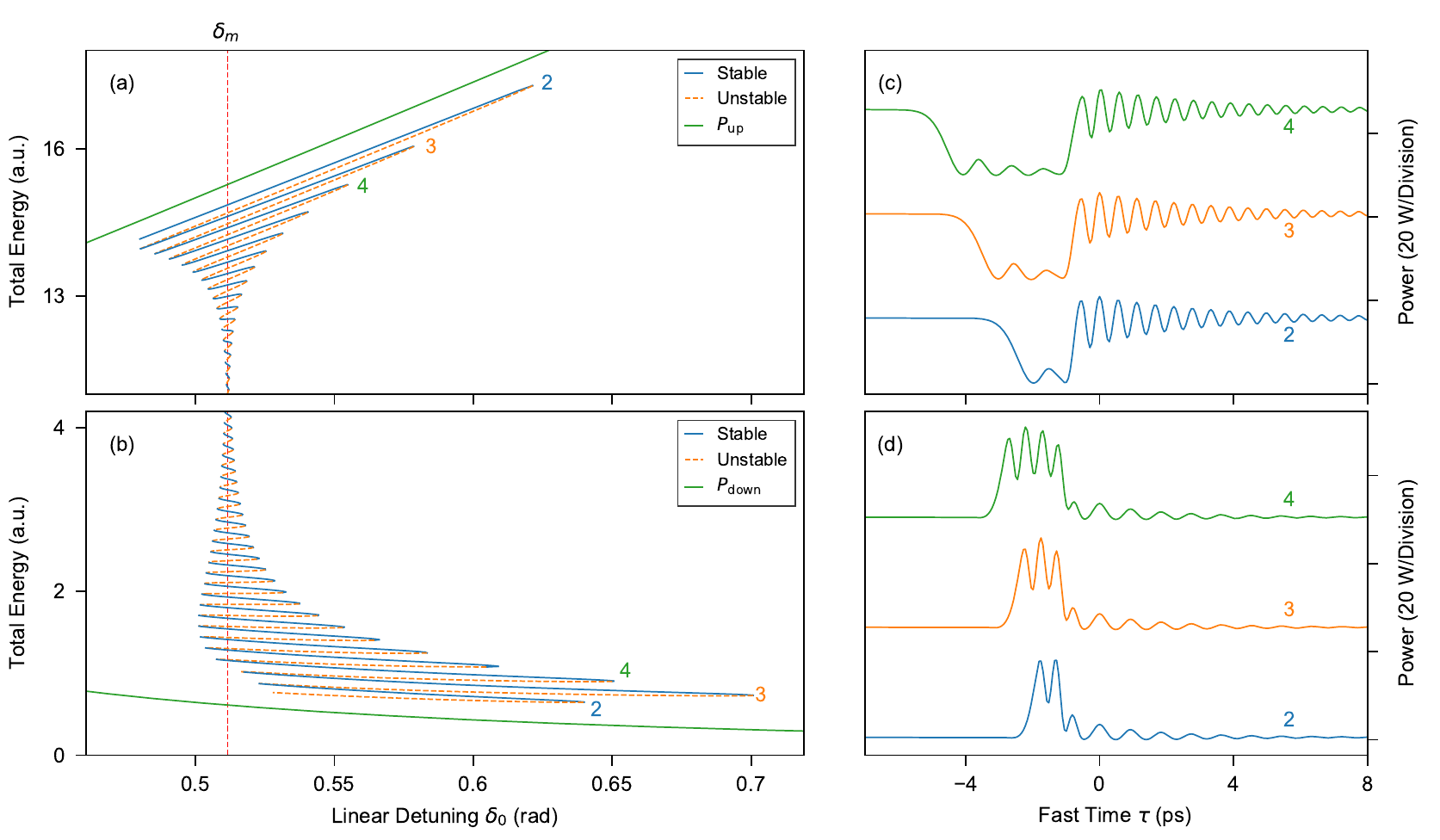}
	\caption{Numerically simulated collapsed snaking bifurcation structure. (a) and (b) show the bifurcation curves of dark and bright solitons under conditions of normal dispersion driving, respectively. Blue solid (dashed orange) curves correspond to stable (unstable) localized structures, whilst the green solid curves correspond to the upper and lower cw solutions of the system. (c) and (d) show examples of temporal profiles corresponding to dark and bright solitons, respectively. The simulation parameters are similar (albeit not identical) to the values used in our experiments: $L = 26~\mathrm{m}$, $\mathcal{F} = 45$, $\theta = 0.05$, $\beta_2 = 0.5~\mathrm{ps^2/km}$, $\beta_3 = 0.2~\mathrm{ps^3/km}$, $\gamma=1.5~\mathrm{W^{-1}km^{-1}}$, $\theta = 0.05$, $P_\mathrm{in} = 2.9~\mathrm{W}$. The calculations ignore fourth-order dispersion and stimulated Raman scattering. }\label{collapsedS}
\end{figure*}

In Figs.~\ref{collapsedS}(a) and (b), we show a typical collapsed snaking bifurcation structure, obtained by finding the localized steady-state solutions of Eq.~\eqref{LEE1} under conditions of normal dispersion driving (see figure caption for parameters) using a Newton-Raphson continuation algorithm. Here we plot the integrated energy of the intracavity field as a function of the cavity detuning, with solid blue (dashed orange) curves corresponding to stable (unstable) solutions. Also shown as green solid curves are the energy levels corresponding to the homogeneous continuous wave (cw) steady-state solutions of the system, which map out an $S$-shaped profile characteristic of bistability.

As can be seen in Figs.~\ref{collapsedS}(a) and (b), the bifurcation curve is composed of layers that fold on top of each other and that overlap over a finite range of detunings. Each layer corresponds to a localized structure characterized by a certain number of dips [for dark solitons (a)] or peaks [for bright solitons, (b)] in its temporal profile; Figs.~\ref{collapsedS}(c) and (d) highlight that, moving vertically from one layer to another causes the number of peaks (or dips) to change by one. It is worth noting that, for parameters used in Fig.~\ref{collapsedS}, the single-peak bright soliton does not exist (see also section ``Bifurcation curve corresponding to experimental parameters'' below).

\begin{figure*}[t!]
	\includegraphics[width=\linewidth]{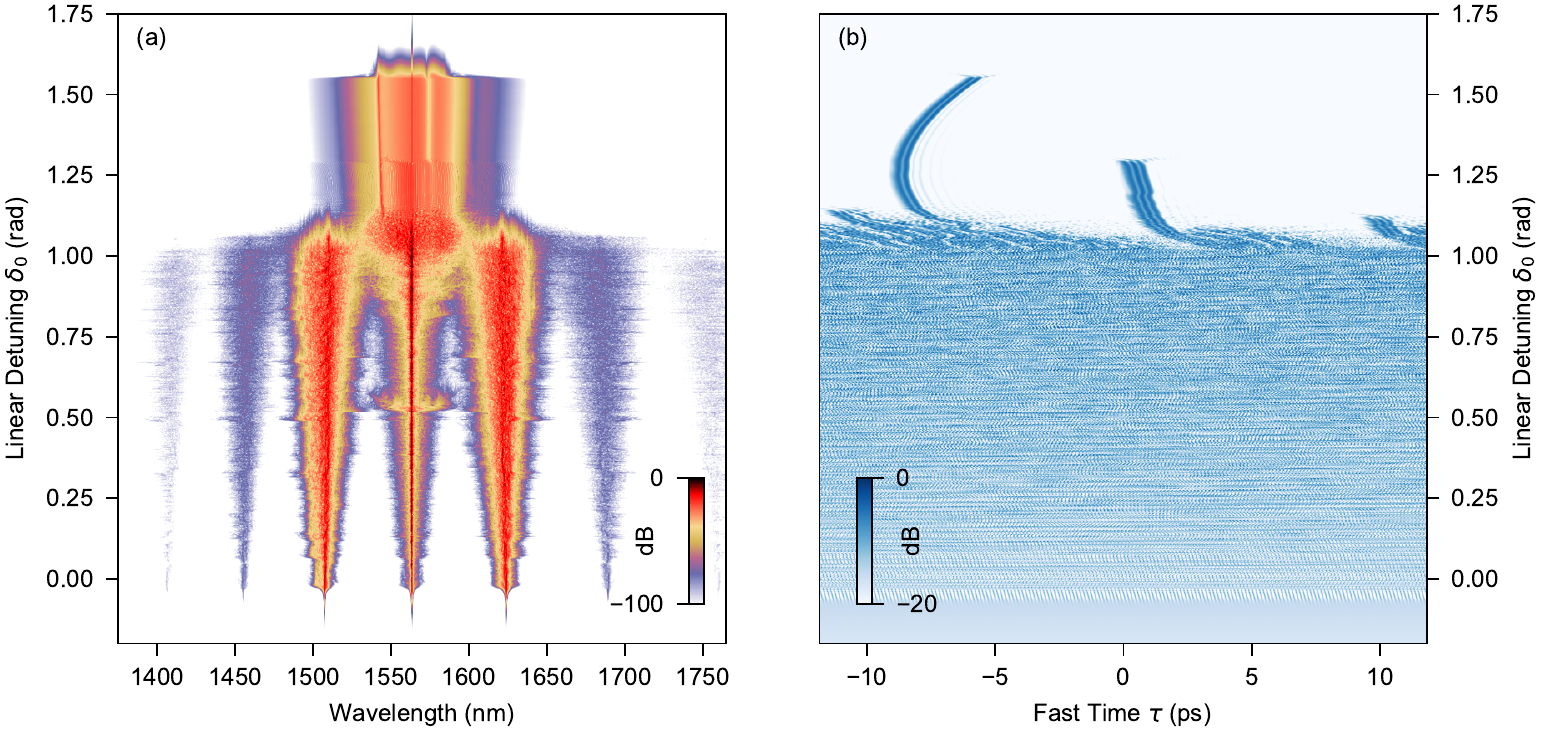}
	\caption{Numerically simulated results, showing the spontaneous excitation of bright localized structures via fourth-order modulation instability. Pseudo-color plots in (a) and (b) respectively show the spectral and temporal evolution of the intracavity field as the detuning is slowly increased. The temporal profiles are plotted in a reference frame that is close to the group velocity of the bright structures. Simulation parameters: $L = 26~\mathrm{m}$, $\mathcal{F} = 45$, $\theta = 0.05$, $\beta_2 = 0.133~\mathrm{ps^2/km}$, $\beta_3 = 0.133~\mathrm{ps^3/km}$, $\beta_4 = -9\times10^{-4}~\mathrm{ps^4/km}$, $\gamma=1.5~\mathrm{W^{-1}km^{-1}}$, $\theta = 0.05$, $P_\mathrm{in} = 6.7~\mathrm{W}$. The calculations ignore stimulated Raman scattering. }\label{b4dynamics}
\end{figure*}

A salient characteristic of the collapsed snaking bifurcation curve is that the range of existence of bright (dark) solitons tends to reduce as the number of peaks (dips) in its temporal profile increases. For very large number of peaks (or dips), the structures only exist over a narrow range around a particular detuning $\delta_\mathrm{M}$. This corresponds to the so-called Maxwell point: a unique point where fronts are motionless even without interacting, thus theoretically allowing dark or bright structures with arbitrary width.

\section{Impact of fourth-order dispersion}

\noindent As mentioned in our main manuscript, the primary impact of fourth-order dispersion is that it renders the upper-state of the bistable cw cavity response unstable against modulation instability (MI). Ignoring SRS, the phase-matching condition for the four-wave mixing process that underpins MI can be written as~\cite{sayson_octave-spanning_2019}
\begin{equation}\label{PM}
\frac{\beta_4\Omega^4L}{24}+ \frac{\beta_2\Omega^2L}{2}  + 2\gamma PL - \delta_0 = 0,
\end{equation}
where $\Omega = \omega-\omega_\mathrm{p}$ describes an (angular) frequency shift of a sideband at $\omega$ from the pump at $\omega_\mathrm{p}$, $P$ is the power level of the cw steady-state solution of Eq.~\eqref{LEE1}, and $\beta_2$ and $\beta_4$ are the second- and fourth-order dispersion coefficients \emph{at the pump frequency}, respectively. It should be evident that a negative fourth-order dispersion coefficient can compensate for positive second-order dispersion at large frequency-shifts, thus enabling phase-matching and MI under conditions of normal dispersion driving.

Our experimental system exhibits a negative fourth-order dispersion coefficient, and so the upper cw state exhibits MI. As a consequence, it is not possible to sustain dark solitons in our system. (Since the cw state that surrounds them is unstable.) With regards to bright solitons, the main impact of fourth-order dispersion is to offer a convenient route to excite the solitons simply by scanning the laser frequency (detuning) across a cavity resonance. This is illustrated in Fig.~\ref{b4dynamics}, where we show an example of numerically simulated dynamics as the detuning is adiabatically scanned over a resonance, obtained by integrating Eq.~\eqref{LEE1} with the split-step Fourier method [see figure caption for parameters]. As can be seen, the intracavity field initially corresponds to a cw state, but then undergoes MI at a detuning of about $\delta_0\approx -0.1~\mathrm{rad}$. As the detuning increases further beyond $\delta_0\approx 1.15~\mathrm{rad}$, we see the spontaneous emergence of bright localized structures characterized by different numbers of peaks that cease to exist at different detunings.

\begin{figure}[b]
	\includegraphics[scale=1]{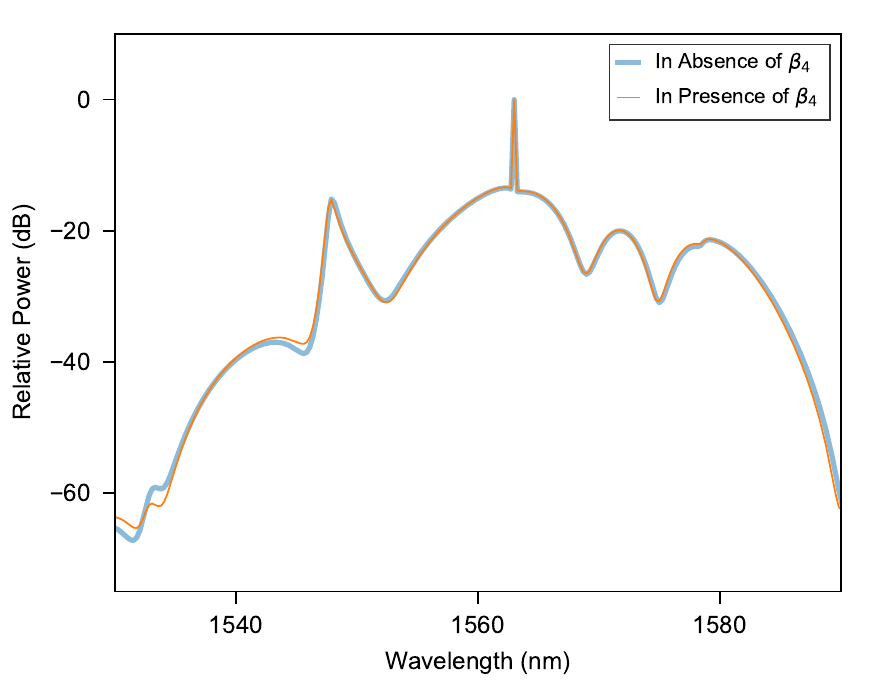}
	\caption{Theoretically predicted soliton spectrum in the presence ($\beta_4 = -9\times10^{-4}~\mathrm{ps^2/km}$, orange curve) and absence ($\beta_4 = 0$, blue curve) of fourth-order dispersion. The linear cavity detuning $\delta_0 = 0.7~\mathrm{rad}$ with all other parameters as in Fig.~\ref{b4dynamics} }\label{pattern or not}
\end{figure}

We must emphasize that, other than facilitating their excitation, fourth-order dispersion does not significantly affect the characteristics of the bright solitons.  This is illustrated in Fig.~\ref{pattern or not}, where we plot spectra corresponding to steady-state soliton solutions of Eq.~\eqref{LEE1} both in the presence and absence of fourth-order dispersion [other parameters as in Fig.~\ref{b4dynamics}]. Two important conclusions can be drawn: (i) the soliton indeed exists even without fourth-order dispersion, and (ii) fourth-order dispersion does not materially impact on the soliton characteristics.  In this context, the fact that the soliton solutions  exist even in the complete absence of fourth-order dispersion should make clear that the solitons \emph{do not} correspond to localized elements of any $\beta_4$-induced MI pattern. This fact is also evidenced by the disparity between spectra in the MI and soliton regimes seen in Fig.~\ref{b4dynamics}(a). It appears that, whilst $\beta_4$ can facilitate the excitation of the normal-dispersion solitons, the origins of the two phenomena are dynamically distinct, one related to third-order dispersion (solitons) and the other related to fourth-order dispersion (MI).

\section{Bifurcation curve corresponding to experimental parameters}

\begin{figure}[b]
	\includegraphics[scale=1]{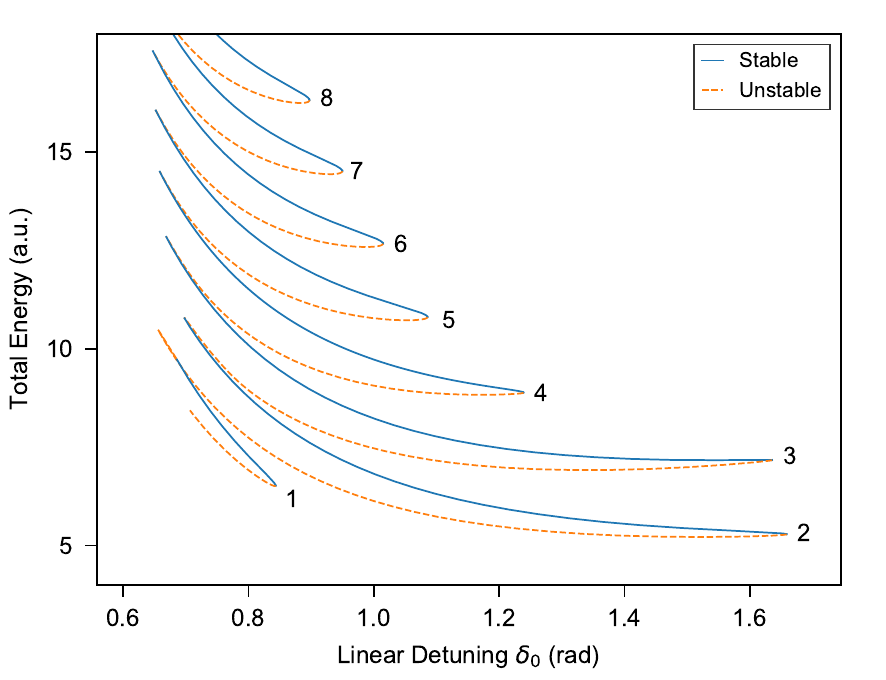}
	\caption{Theoretically predicted bifurcation curve of bright solitons corresponding to our experimental parameters. Numbers correspond to the number of peaks in the solitons' temporal profile. The curves are somewhat different to those shown in Fig. 3 of our main manuscript as the total energy here includes the cw pump component.}\label{exprbif}
\end{figure}

\noindent For completeness, Fig.~\ref{exprbif} shows the numerically computed bifurcation curve of bright solitons for our experimental parameters. (The curve is obtained from Eq.~\eqref{LEE1} with both higher-order dispersion and stimulated Raman scattering included.) Theoretical results shown in Figs.~2--4 of our main manuscript are representative of this bifurcation curve.

It is worth noting that the bifurcation curve shown in Fig.~\ref{exprbif} includes the single-peak soliton that is reminiscent of conventional cavity solitons in the anomalous dispersion regime. This should be contrasted with the illustrative results shown in Fig.~\ref{collapsedS}: the parameters used in that case are such that single-peak solitons do not exist. Taken together, these results illustrate that (i)~multi-peak bright structures can exist even for parameters for which single-peak solitons do not exist, and (ii)~when single-peak solitons do exist, they are part of the same bifurcation curve as the multi-peak structures.

\section{measurement of resonator dispersion}

\begin{figure}[b]
	\includegraphics[scale=1]{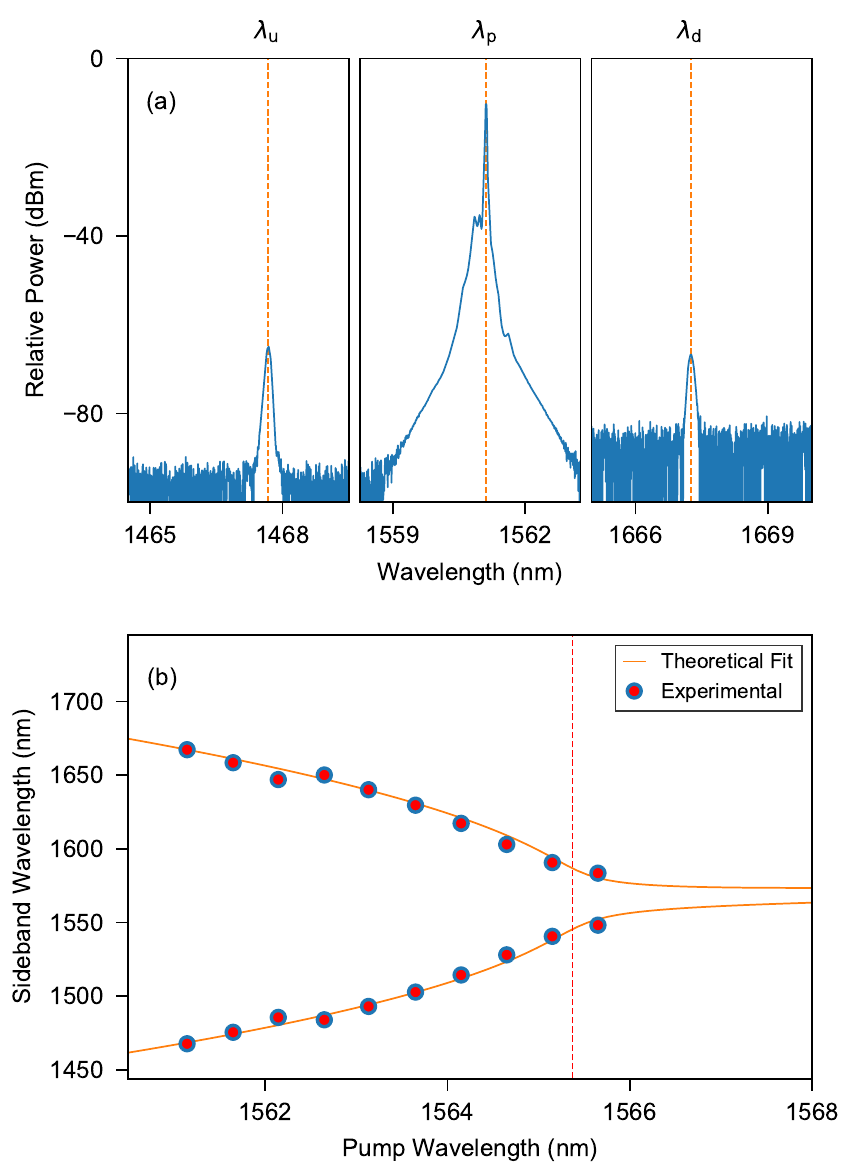}
	\caption{(a) Example of an MI spectrum recorded with driving wavelength $\lambda_\mathrm{P} = 1561$~nm. The entire spectrum is divided into three panels for clarity: one for each sideband (left and right) and one for the pump (middle). (b) Solid dots show experimentally measured sideband wavelengths for a range of pump wavelengths. Solid curves show theoretical fit as explained in the text. The location of the zero-dispersion wavelength is indicated by the dashed red line.}\label{dispersion measurement}
\end{figure}

\noindent To measure the dispersion of the resonator used in our experiments, we record the MI sidebands at various pump wavelengths, and fit Eq.~\eqref{PM} to the experimentally observed tuning curve. To facilitate the analysis of the results, the cavity detuning is stabilized at zero, and the input power is kept as low as possible (whilst still observing MI). Thanks to these procedures, the two last terms in Eq.~\eqref{PM} can be ignored.

Figure~\ref{dispersion measurement}(a) shows an illustrative spectrum measured for a pump wavelength of $\lambda_\mathrm{P} = 1561$~nm, which lies in the normal dispersion regime. We observe widely detuned sidebands at $\lambda_\mathrm{u} = 1467.68$ and $\lambda_\mathrm{d} = 1667.26$~nm. By measuring such spectra for a range of pump wavelengths, we obtain the tuning graph shown by the solid circles in Fig.~\ref{dispersion measurement}(b). We then extract the dispersion characteristics of the resonator by fitting Eq.~\eqref{PM} into the measured data. More specifically, there are three fitting parameters: the zero-dispersion wavelength $\lambda_\mathrm{ZDW}$ and the third- and fourth-order dispersion coefficients $\beta_{3,\mathrm{ZDW}}$ and $\beta_{4,\mathrm{ZDW}}$ at the ZDW, respectively. These parameters define the dispersion coefficients $\beta_2$ and $\beta_4$ appearing in Eq.~\eqref{PM} at different pump wavelengths $\lambda_\mathrm{P}$ viz.
\begin{align*}
\beta_2 &= \beta_{3,\text{ZDW}}\Omega + \frac{\beta_{4,\text{ZDW}}}{2}\Omega^2, \\
\beta_4 &= \beta_{4,\text{ZDW}},
\end{align*}
where $\Omega = 2\pi c\left[\lambda_\mathrm{P}^{-1} - \lambda_\mathrm{ZDW}^{-1} \right]$ with $c$ the speed of light in vacuum.

\begin{figure}[t]
	\includegraphics[scale=1]{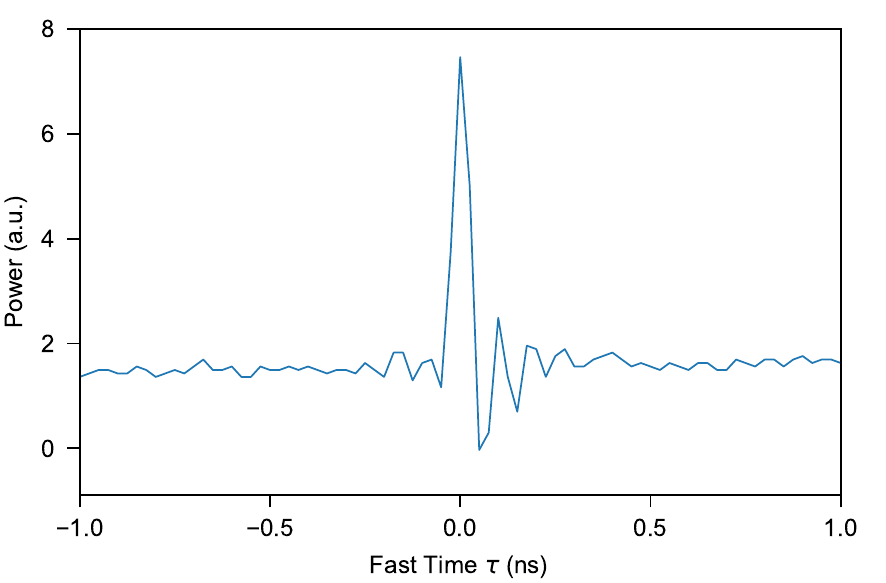}
	\caption{Typical photodetector trace measured at the 1\% output port of our cavity, providing direct evidence of a bright localized structure.}\label{bright}
\end{figure}

One caveat with the method described above is that, due to the presence of three unknowns, an accurate fit to the equation using sideband data alone is relatively difficult, especially if the driving wavelength range is limited. In particular, we find that multiple combinations of parameters can yield reasonably good agreement with the experimental data. In order to reduce ambiguity, in addition to fitting Eq.~\ref{PM}, we also compare the experimentally measured spectra of normal-dispersion solitons with the spectra predicted by the theory for the dispersion coefficients that provide a reasonable fit to the phase matching curve. This two-step procedure yields the following parameters:  $\lambda_\mathrm{ZDW} = 1565.37$~nm, $\beta_{2,\mathrm{ZDW}} = 0~\mathrm{ps}^2/\mathrm{km}$ (by definition), $\beta_{3,\mathrm{ZDW}} = 0.135~\mathrm{ps}^2/\mathrm{km}$ and $\beta_{4,\mathrm{ZDW}} = -9\times10^{-4}~\mathrm{ps}^4/\mathrm{km}$. The solid curves in Fig.~\ref{dispersion measurement} show the corresponding fit, and we see very good agreement. We believe these dispersion parameters obtained are very accurate, as they are capable of producing theoretical predictions of normal-dispersion solitons that agree with our experimental observations across a range of wavelengths, pump powers, linear detuning, and soliton types (as demonstrated in the main text).

\section{Evidence of bright structures}
\noindent All of the localized structures studied in our work correspond to \emph{bright} structures, i.e., pulses of light that sit atop a low-intensity background. This fact can be observed directly from the photodetector signals measured at the 1\% output port of our cavity. Figure~\ref{bright} shows a typical example of such  a photodetector signal, and indeed, we see a pulse that sits atop a background. We must emphasize that the 80~ps response time of our detection system is far too slow to resolve the soliton's temporal profile, yet nevertheless, the signal shown in Fig.~\ref{bright} provides unequivocal evidence of the structure's bright (rather than dark) character.